\documentclass[epj,nopacs]{svjour2}
\usepackage{epsfig,pst-plot,colordvi}
\usepackage{graphicx}
\usepackage{epsf,amssymb}
\newcommand{\beq}{\begin{equation}}
\newcommand{\eeq}{\end{equation}}
\newcommand{\bi}{\begin{itemize}}
\newcommand{\ei}{\end{itemize}}
\newcommand{\beqar}{\begin{eqnarray}}
\newcommand{\eeqar}{\end{eqnarray}}
\newcommand{\beann}{\begin{eqnarray*}}
\newcommand{\eeann}{\end{eqnarray*}}

\newcommand{\nn}{\nonumber}

\newcommand{\bpi}{\beta_\pi}

\newcommand{\bea}{\begin{eqnarray}}
\newcommand{\eea}{\end{eqnarray}}

\newcommand{\be}{\begin{equation}}
\newcommand{\ba}{\begin{eqnarray}}
\newcommand{\ee}{\end{equation}}
\newcommand{\ea}{\end{eqnarray}}

\newcommand{\tr}{\mbox{tr}}

\newcommand{\dilog}{\mbox{Li}_2}
\newcommand{\GeV}{\mbox{GeV}}

\newcommand{\ol}{\overline}

\newcommand{\dafne}{DA$\Phi$NE }

\newcommand{\epmppm}{e^+ e^- \rightarrow \pi^+ \pi^- }

\newcommand{\amuh}{a_{\mu}^{\rm had}}

\newcommand{\lapprox}{\raisebox{-.2ex}{$\stackrel{\textstyle<}{\raisebox{-.6ex}[0ex][0ex]{$\sim$}}$}}

\let\epsilon\varepsilon

\begin{document}
\thispagestyle{empty}
\title{
\begin{flushright}{\normalsize \rm
DESY 02-155
\vspace*{0.5cm}}
\end{flushright}
\begin{center}
{Measuring the FSR--inclusive $\pi^+\pi^-$ cross
section}\thanks{Work supported in part by TMR, EC-Contract
No.~ERBFMRX-CT980169 (EURODA$\Phi$NE), EC-Contract
No.~HPRN-CT-2000-00149 (Physics at Colliders), TARI project HPRI-CT-1999-0008
and by the Alexander von Humboldt Stiftung.
} \end{center}}
\author{J.~Gluza\inst{1,3}, A.~Hoefer\inst{2}, S.~Jadach\inst{2},
F.~Jegerlehner\inst{3}}
\institute {Institute of Physics, University of
Silesia, Uniwersytecka 4, 40-007 Katowice, Poland
\and Institute of Nuclear Physics, Radzikowskiego 152, 31-342 Krakow, Poland
\and DESY Zeuthen, Platanenallee 6, D-15738 Zeuthen, Germany}
\abstract{Final state radiation (FSR) in pion--pair production cannot be
calculated reliably because of the composite structure of the pions.
However, FSR corrections have to be taken into account for a precise
evaluation of the hadronic contribution to $g-2$ of the muon. The role
of FSR in both energy scan and radiative return experiments is
discussed. It is shown how FSR influences the pion form factor
extraction from experimental data and, as a consequence, the
evaluation of $a_{\mu}^{\rm had}$. In fact the $O(\alpha)$ FSR
corrections should be included to reach the precision we are aiming
at. We argue that for an extraction of the desired FSR--inclusive
cross section $\sigma^{(\gamma)}_{\rm had}$ a photon--inclusive scan
measurement of the ``$e^+e^- \to \pi^+\pi^- + {\rm photons}$'' cross
section is needed. For exclusive scan and radiative return
measurements in contrast we have to rely on ad hoc FSR models if we
want to obtain either $\sigma^{(\gamma)}_{\rm had}$ or the
FSR--exclusive cross section $\sigma^{(0)}_{\rm had}$. We thus
advocate to consider seriously precise photon--inclusive energy scan
measurements at present and future low energy $e^+e^-$--facilities.
Then together with radiative return measurements from DA$\Phi$NE and
BABAR and forthcoming scan measurements at VEPP-2000 we have a good
chance to substantially improve the evaluation of $a_{\mu}^{\rm had}$
in the future.}

\titlerunning{Measuring the FSR inclusive ...}
\authorrunning{J.~Gluza, A.~Hoefer, S.~Jadach, F.~Jegerlehner}
\maketitle

\section{Introduction}
Photon vacuum polarization effects are sizable and therefore play an
important role in electroweak precision physics. Because of the strong
interactions between quarks and gluons the contributions of the low
energy hadrons cannot be calculated by perturbative QCD. However, they
may be obtained via a dispersion integral over the experimental
$e^+e^-$ annihilation data. A precise evaluation of hadronic effects
in quantities like the running fine structure constant $\alpha(s)$ and
of the muon anomalous magnetic moment $a_\mu$ thus depends directly on
the precision of low energy ``$e^+e^- \to {\rm hadrons}$'' cross sections
$\sigma_{\rm had}$~\cite{r1,Alemany:1997tn,Davier:2002dy}. Further
theoretical efforts may help to some extent to reduce the theoretical
uncertainties of these quantities~\cite{Heiri}. However, new
measurements of $\sigma_{\rm had}$ are indispensable for achieving
substantial progress. Indeed, remarkable improvements have been
achieved in recent years by the Collaborations
CMD-2~\cite{Akhmetshin:2001ig} at Novosibirsk and
BES-II~\cite{Bai:2001ct} at Beijing. New results are expected soon
from radiative return experiments by
KLOE~\cite{dafne,Valeriani:2002yk} at the $\Phi$--factory DA$\Phi$NE
at Frascati and from the $B$--factory at SLAC with
BABAR~\cite{Solodov:2002xu}.

The muon g-2 experiment at Brookhaven now has reached the level of
0.7~ppm in precision~\cite{Brown:2001mg,Bennett:2002jb} for a
measurement of $a_\mu$ and depending on which evaluation of
$a_\mu^{\rm had}$ is adopted~\cite{Davier:2002dy} reveals a deviation
from the Standard Model prediction
which could be as large as 3 standard deviations. Since the main
source of uncertainty of the SM prediction arises from the hadronic
contributions, a careful reconsideration of the determination of
$a_\mu^{\rm had}$ is mandatory. In fact existing low energy ``$e^+e^-
\to \pi^+\pi^-$'' data are inconsistent with the corresponding $I=1$
part obtained via CVC (conserved iso-vector current) from hadronic
$\tau$--decay spectra~\cite{Davier:2002dy,Eidelman:1994qm}. This is a
problem which most likely can only be resolved by new
experiments. Needless to say that the experimental inconsistencies
also reduce our possibilities to obtain a more precise determination
of $a_\mu^{\rm had}$ and hence to draw conclusions about possible
``new physics'' which also would contribute to $a_\mu$.

Experiments that measure $\sigma_{\rm had}$ do this either via an
energy scan ($2m_\pi\leq \sqrt{s} \leq E_{\rm max}$) or they measure
the invariant mass distribution of the hadronic final states
$d\sigma_{\rm had}/ds'$ ($s' \leq s$) at meson factories running at
fixed $s$, using the radiative return due to the emission of hard
initial state photons. From $d\sigma_{\rm had}/ds'$ the cross section
$\sigma_{\rm had}(s')$ can here be extracted by factoring out the
photon radiation\footnote{As has been pointed out
in~\cite{Hoefer:2001mx} already, the radiative return ``mechanism'' at
leading order has the nice property that the usual convolution
integral, relating the observed cross section (which includes photon
radiation effects) to the physical cross section of actual interest,
appears de-convoluted (photon radiation acts as a spectral analyzer)
such that instead of factorization under convolution integrals one has
point by point factorization. Higher order effects which give raise to
multiple convolution integrals of course spoil this simple picture
since by taking one derivative we get ride of one integration only.}.

At increasing precision it becomes more and more difficult and
challenging to extract the relevant ``pseudo observable'' quantities
with adequate precision. By ``pseudo observable'' one understands
quantities obtained from raw experimental data only via some
theoretical input. For example, one has to unfold the raw data from
photon radiation effects, where the initial state radiation (ISR) is
universal to all $e^+e^-$--annihilation processes, while the final
state radiation (FSR) and the initial--final state interference (IFS)
are process specific. The ``pseudo observable'' we are interested in
is the ``hadronic blob'' which corresponds to the imaginary part of
the correlator of two hadronic electromagnetic currents: the
one--photon irreducible contributions to the photon vacuum
polarization (see~\cite{Hoefer:2001mx} and references therein). Here
we concentrate on low energy $\pi$--pair production, a relatively
simple hadron production channel which is dominating the hadronic
contribution to the muon g-2.

For both the scan and the radiative return method we are facing three
major sources of uncertainty affecting the extraction of
$\sigma_{\pi\pi}$: the experimental error, the theoretical error due
to neglecting higher order QED corrections and finally the uncertainty
related to non-perturbative effects related to photon radiation from
the final hadronic state. Currently, great efforts are made to reduce
the experimental error below the one per cent
level~\cite{Akhmetshin:2001ig,dafne,Valeriani:2002yk,Solodov:2002xu}.
QED corrections concerning low energy pion pair production have been
considered
e.g.~in~\cite{Hoefer:2001mx,Arbuzov:1997je,karlsruhe,Arbuzov:1998te}. In
the present article we will focus on the last of the mentioned error
sources, the model error related to photons radiated from the final
hadronic state. Here the problem is that the radiation of photons by
the pions is poorly understood theoretically. Since perturbative QCD
breaks down at low energies it is not possible to treat the final
state pions in terms of their constituent quarks. On the other hand
hard photons participating in the scattering process do probe the pion
sub--structure. Treating pions as point--like scalar particles by
simply applying scalar QED (sQED) is therefore also not a solution to
the problem. What makes things even more complicated is the fact that
we have to deal with non--perturbative QCD effects like intermediate
$\rho$ or $\omega$ resonances and photon radiation from such a
hadronic state cannot be treated in a straightforward way. However,
this contribution of real photon emission can be expected to be less
important than the radiation from the final state pions. This is
because the net charge of the intermediate hadronic state is zero and
in addition the de Broglie wavelength of the dominant $\rho$ and
$\omega$ resonances is relatively small in respect to the typical
wavelength of the radiated photons. Only sufficiently hard photons are
able to probe the sub--structure of a hadronic composite state. The
pions on the other hand are charged and have a much longer de Broglie
wavelength. Unfortunately a similar argument does not help for the
virtual corrections since virtual hard photons are always included and
also cannot be eliminated by cutting out the ``trouble--making'' part
of the phase space as it is possible for real photons. As a
consequence their magnitude is not known and they cannot be subtracted
from the hadronic final state without relying on specific models like
sQED.

As mentioned above the quantity of interest is the correlator of the
hadronic component of two electromagnetic currents including strong as
well as electroweak corrections. From the latter only the photonic
corrections are sizable. They correspond to the FSR correction in the
hadron production
processes~\cite{Melnikov:2001uw,DeTroconiz:2001wt}. Since these
corrections cannot be calculated reliably, the aim is to measure, if
possible, the hadronic cross section $\sigma_{\rm had}^{(\gamma)}$
that is dressed by final state photons. We would obtain in this way
directly the quantity to be inserted into the dispersion integrals for
the hadronic contribution to the running fine structure constant
$\alpha(s)$ and to the muon anomalous magnetic moment $a_{\mu}$,
respectively, at the next to leading level of accuracy.

Including FSR means to include photonic
corrections to the irreducible hadronic photon self--energy. We thus
address the question whether we can circumvent the FSR problem by
performing an inclusive measurement, i.e., undress the data from
ISR only. The question has been discussed already
in a previous paper for the radiative return
scenario~\cite{Hoefer:2001mx}. The result was that in this case an
inclusive measurement does not yield the quantity of interest at
sufficient precision. In other words, without substantial loss of
precision one cannot avoid the necessity to undress from all photon
radiation, including the complete treatment of FSR. Lacking a precise
theoretical understanding of photon radiation by hadrons, however, one
has to rely on model assumptions like ``generalized sQED'' for
the pions (treating pions as point--like modulo a form factor)
to extract the undressed (FSR-exclusive) cross section 
$\sigma^{(0)}_{\rm had}(s)$ from the
data in a first step. In order to obtain the FSR--inclusive
cross section $\sigma^{(\gamma)}_{\rm had}(s)$ one has to add 
the appropriate FSR contribution ``by hand'' at the end.

On the other hand for completely inclusive scan measurements we can
use the fact that ISR and FSR factorize to ``subtract'' ISR from the
observed inclusive total cross section $\sigma_{\rm obs}$, leaving, up
to $O(\alpha^2)$ IFS contributions, the desired FSR--inclusive
cross section $\sigma^{(\gamma)}_{\rm had}(s)$. As we will see for pion
pair production such a ``subtraction'' of ISR is also possible with
excellent precision for realistic cuts on the pion angles provided
they are chosen such that they
break the ISR$\otimes$FSR-factorization only slightly. The inclusive
measurement requires a high quality detector with high acceptance and
good separation of $\pi^0$ vs. $\gamma$ ($\pi^+\pi^-\pi^0$
background).

At CMD-2~\cite{Akhmetshin:2001ig} so far a different strategy has been
used which we will call the exclusive scan measurement. Here an event
selection is applied such that only soft real photons are included
which then can be corrected away. While there are no problems with
real hard photons in this case one still has the problem that the
virtual photon contributions from the loops include hard photon
effects. The virtual contributions must be subtracted as well as one
has to apply the Bloch--Nordsieck construction in order to get an
infrared--finite cross section. Because the effective theory applied
(generalized sQED) is renormalizable one obtains infrared (IR) 
and ultraviolet (UV) finite results.

Note that existing data at present do not allow us to determine
$\sigma^{(\gamma)}_{\rm had}(s)$ as required for a precise determination
of its contribution to $\amuh$. Modeling FSR by sQED we may
estimate the size of the effect we have in mind: it is given at
leading perturbative order by $\delta^{\gamma} \amuh=(38.6 \pm 1.0)
\times 10^{-11}$. This has to be confronted with the final precision
$\delta^{\rm exp} \amuh \sim 40
\times 10^{-11}$ expected from the Brookhaven muon g-2 experiment.

Before settling this 1 $\sigma$ (in terms of the expected final
experimental precision) effect, it is urgent to clarify the origin of
the 3 times larger discrepancy between $e^+e^- \to \pi^+\pi^-$ data
and corresponding data obtained via CVC from $\tau$
spectral--functions (in the energy range just above the
$\rho$--resonance), and the present unclear status of the $\rho$ mass
and width~\cite{Hagiwara:pw}. This issue can certainly be settled by
the radiative return experiments with KLOE~\cite{Valeriani:2002yk} at
LNF/Frascati and with BABAR~\cite{Solodov:2002xu} at SLAC~\footnote{We
should mention that another very problematic energy region exists
where experimental data are very poor or even controversial and which
is important for the precise evaluation of $\amuh$: the energy range
1.4 to 2.0 GeV (between the upper limit of the VEPP-2M machine at
Novosibirsk and the lower limit of the BEPC machine at Beijing).
Radiative return measurements with BABAR and results expected from
VEPP-2000 (upgraded VEPP-2M) will substantially improve results in
this range.}. However, a new energy scan experiment, which is anyway
mandatory for a clean measurement of the FSR--inclusive cross section,
could also help to clarify the origin of the observed deviations. In
addition, as we shall argue below, in radiative return measurements in
order to get rid of the model dependent FSR contribution, at least one
of the following conditions has to be fulfilled: i)
$\sigma_{\pi\pi}(s'<s) \gg \sigma_{\pi\pi}(s)$ (true especially for
the $\rho$ resonance region); ii) $s' \simeq s$ (soft photon region);
iii) suppression of FSR by kinematic cuts. As we will see, at
$\phi$--factories model dependence becomes an insurmountable problem
at low energies below about 500 MeV where we have to deal with large
contributions of hard FSR photons which cannot be suppressed by cuts.

The above remarks together with the results presented in this paper
strongly suggest that it would be desirable to revitalize the idea
to perform an energy scan at the DA$\Phi$NE machine at
Frascati~\cite{Maiani:ve} in a second step after running as a
$\Phi$--factory.

In the next section we discuss the model error of pion form factor
extraction connected to the radiation of photons
from hadronic final states in inclusive and exclusive
scan scenarios.
In addition we analyze to what extent kinematic cuts on
the pion angles change the picture
and, for the inclusive scan scenario, consider the impact of the model
uncertainty on the determination of $a_\mu^{\rm had}$.
Section 3 is devoted to the model uncertainty of extracting
the pion form factor $|F_\pi(s')|^2$ for fixed $s$ 
in radiative return experiments.
We also address the question if the FSR contribution and its
related model error can be estimated from a measurement of
the pion forward-backward asymmetry $A_{FB}$.
In Appendix~A we derive a general, model independent formula
for the inclusive cross section
$\sigma(e^+e^-\to\gamma^*\to X+ \rm photons)$,
where ISR and FSR are treated in a factorized form, X being an arbitrary
non--photonic final state.
In Appendix~B some of the used formulas connected to FSR within
sQED or fermionic QED (fQED) are collected.
\section{Model errors for inclusive and exclusive measurements
in scan experiments}
\subsection{Inclusive scenario}
We first present a case study of ``$e^+e^- \to \pi^+\pi^- \:+\: n 
\;\gamma$'' ($n=0,1,2,\ldots$).
Experimentally on an event by event basis it is not possible to
distinguish a final state from an initial state photon. In an
inclusive measurement events with any number of (initial and/or
final state) photons are counted. The major question will be to what
extent and at what accuracy we may evaluate FSR--inclusive cross sections
from the experimental data. We first consider the measurement of the
FSR--inclusive cross section\footnote{In the following we drop the labels
``had'' or ``$\pi \pi$'' for cross sections like $\sigma^{(0)}$ and
$\sigma^{(\gamma)}$.} $\sigma^{(\gamma)}(s)$ in energy
scan experiments.

Suppose for the moment, that we would be able to calculate photon
radiation from pions. Then we would have two possibilities: 
\begin{itemize}
\item[(i)]
determine the undressed cross section $\sigma^{(0)}(s)$ by unfolding the
observed cross section from all photon radiation and add the FSR as
calculated by perturbation theory with desired accuracy, which
yields $\sigma^{(\gamma)}$; 
\item[(ii)]
 determine an
FSR--inclusive cross section by unfolding only the calculated ISR from
the observed cross section, which yields $\hat{\sigma}^{(\gamma)}(s)$.
\end{itemize}
The question then is to what accuracy does
$\hat{\sigma}^{(\gamma)}(s)$ approximate
$\sigma^{(\gamma)}(s)$. Since, actually, we do not know how to
calculate photon radiation from pions in a model independent way only
the second approach is able to give a model-independent answer,
however, then we do not know how well $\hat{\sigma}^{(\gamma)}(s)$
approximates the quantity of interest $\sigma^{(\gamma)}(s)$. What we
will do in this case is a ``guesstimate'' of the quality of the
approximation by modeling FSR by generalized sQED.

For the error estimate the following factorization theorem is crucial:
Neglecting the IFS contribution, being of $O(\alpha^2)$
due to charge conjugation invariance, the inclusive total cross section
$\sigma_{\rm obs}(s)$ may be written in a factorized form,
\ba
\sigma_{\rm obs}(s) &=&
\int ds_V\;\sigma^{(\gamma)}(s_V)\;\rho_{\rm ini}^{\rm
incl}(s,s_V) +O(\alpha^2)_{\rm IFS} \;, \nonumber \\
\label{main}
\ea
which means that FSR and ISR can be treated independently from each
other. Details are given in Appendix~A [see (\ref{inclusive}),
(\ref{inclusivealpha})]. 
It is worth to stress that the powerful identity (\ref{main})
is not easily recognizable to fixed perturbative order.
Note that (\ref{main}) is quite general once IFS is neglected.
It is based on the fact that we have a neutral current process 
for which we can apply a separation into 
Lorentz--covariant and individually gauge invariant 
initial and final state contributions.
Qualitatively the result may be understood as follows: by the fact that at
low energies the single virtual photon exchange ($1/s$--enhancement)
highly dominates the ``$e^+e^-
\to \pi^+\pi^- \:+\: {\rm n}\: \gamma$'' cross section and due to
the suppression of the IFS (see later) it makes sense to consider 
the process in an
approximation of an  $s$-channel single photon exchange (i.e. diagrams
which factorize into two disconnected parts upon cutting the photon
line). This virtual photon then carries the invariant mass $\sqrt{s_V}$ and
the above convolution is exact up to the indicated missing IFS
effects. We would like to remind the reader that the representation of
the $\pi\pi$--production cross section in terms of the pion form
factor\footnote{Note that $\sigma^{(0)}(s)$ and equivalently
$|F_{\pi}^{(0)}(s)|^2$ are not measurable quantities, as we shall
discuss below. They are useful, theoretically motivated concepts defined in a
world where the electroweak interactions are switched off. In reality
we cannot switch off QED effects and this is part of the problem we
are dealing with in this paper.  If one could calculate
$|F_{\pi}^{(0)}(s)|^2$ for time--like $s$ non-perturbatively in
lattice QCD, this is the quantity what one would take from lattice
QCD.}
\begin{equation}
\sigma^{(0)}(s) = |F_{\pi}^{(0)}(s)|^2\;
\sigma^{\rm 0,point}(s)\;,
\label{ff0}
\end{equation}
with [$\beta_\pi=(1-4m_\pi^2/s)^{1/2}$ is the pion velocity]
\beann
\sigma^{\rm 0,point}(s) =
\frac{\pi}{3}\frac{\alpha^2 \beta^3_\pi}{s}\;,
\eeann
also makes sense strictly only for the one--photon exchange
approximation\footnote{Since in (\ref{ff0}) $F^{(0)}_\pi(s)$ does not
depend on the pion production angle a similar formula (\ref{cut0}) is
valid for the case of angular cuts with the same function
$F^{(0)}_\pi(s)$.}. In this approximation $s_V$ can be neatly
identified with $s$ in $|F_\pi(s)|^2$, in spite of the fact that
$s_V$, as the squared invariant mass of a virtual state, is not an
observable.  Thus in (\ref{main}) $s_V$ is only a formal (unphysical)
integration variable where the boundaries are physical observables:
$4m_\pi^2\leq s_V \leq s$. Nevertheless, we can always fit the pseudo
observable $\sigma^{(\gamma)}(s_V)$ to the observed data $\sigma_{\rm
obs}(s)$ by using (\ref{main}) and thereby determine
$\sigma^{(\gamma)}(s_V)$. The accuracy with which this can be
achieved, up to IFS contributions, is limited by our knowledge of the
initial state radiator function $\rho_{\rm ini}^{\rm incl}(s,s_V)$
only. The latter can be calculated without any model dependence within
perturbative QED (see e.g.~\cite{Jadach}). Since IFS effects are of
$O(\alpha^2)$, the model dependence for the extraction of the
FSR--inclusive cross section is determined by an (as yet unknown)
$O(\alpha^2)$ IFS contribution\footnote{The $O({\alpha^2})$ IFS
is complete for the real photon emission. However, some virtual
contributions are not yet calculated.}.  What is very important is
that IFS interference does not include contributions from leading
logarithms of the kind $\log(s/m_e^2)$.  We may estimate the
$O({\alpha^2})$ effect to be at the per mill level.

As we have already stressed,
our ``master formula'' (\ref{main}) cannot be directly
applied to a real experiment with some cuts and/or detector inefficiencies
(the leading uncertainties are due to the need of extrapolation to the blind 
zones of the measurement).
In a real experiment the influence of these effects will be taken 
into account using a realistic Monte Carlo event generator which features a 
high quality ISR matrix element and some modeling of FSR.
Let us focus therefore on a situation where angular cuts are
present. Then, ISR and FSR phase space integrations cannot be
disentangled as it is possible without cuts [see (\ref{ISFSfact})] and
ISR$\otimes$FSR factorization breaks down at the $O(\alpha)$ level. As a
consequence, to subtract only ISR from the data we have to rely on
specific FSR models. To be able to extract $\sigma^{(\gamma)}(s_V)$
without significant model dependence the condition that {\em the applied
cuts break} ISR$\otimes$FSR {\em factorization only slightly} has to be 
fulfilled.
Here we will investigate the breaking of ISR$\otimes$FSR factorization by
some semi--realistic C--symmetric cuts, $\Theta_{\pi_\pm} \geq \Theta^M_\pi$,
$\Theta_{\pi_\pm}$ being the laboratory angle between the pion momenta
and the beam axis, treating FSR by sQED. For such cuts we then
can write the observed cross section as (for details see Appendix~B
and~\cite{Hoefer:2001mx}; $\Lambda$ is the soft photon energy separating
soft from hard photons)
\ba
\sigma_{\rm obs}^{\rm cut}(s) &=& \sigma_{\rm cut}^{(\gamma)}(s)
\left[1+\delta_{\rm ini}(s,\Lambda)\right] \nn \\
&+&\int_{4m_\pi^2}^{s-2\sqrt{s}\Lambda} ds_V \;
\sigma_{\rm cut}^{(\gamma)}(s_V)\;\rho_{\rm ini}^{\rm cut}(s,s_V)
-\delta_{\rm cut}^{\rm scan}(s) \;, \nn\\ \label{scincl}
\ea
with $\delta_{\rm ini}(s,\Lambda)$ corresponding to the soft plus virtual
and $\rho_{\rm ini}^{\rm cut}(s,s_V)$ corresponding to hard
photon initial state QED corrections.
The $O(\alpha)$ ISR$\otimes$FSR factorization breaking term
$\delta_{\rm cut}^{\rm scan}(s)$ accounts for the missing pion events
which cannot be seen in the experiment:
\ba
\delta_{\rm cut}^{\rm scan}(s) &=& \sigma_{\rm cut}^{(0)}(s)\;
\frac{\alpha}{\pi}\;\left\{\eta(s)
-\eta_{\rm cut}(s)\right\} +O(\alpha^2). \label{defeta}
\ea
$\delta_{\rm cut}^{\rm scan}$
vanishes for the case without cuts (restoration of factorization).
Remember that for C-symmetric angular cuts the $O(\alpha)$ IFS
contribution drops out.

In a world with point-like pions we could calculate
$\delta_{\rm cut}^{\rm scan}(s)$
perturbatively in sQED, where
\ba
\eta_{\rm (cut)}(s) &=& \frac{\pi}{\alpha}\;
\left[\delta_{\rm fin}(s,\Lambda)+
\int_{4m_\pi^2}^{s-2\sqrt{s}\Lambda} ds'\;\rho_{\rm fin}^{\rm (cut)}(s,s')
\right] \;, \nn\\ \label{eta}
\ea
with $s'$ being the square of the invariant mass of the pion pair and
$\delta_{\rm fin}(s,\Lambda)$ and $\rho_{\rm fin}^{\rm (cut)}(s,s')$ the
corresponding FSR corrections given in Appendix~B. For real world
pions we may estimate this term assuming generalized sQED which
at least treats the soft photon part correctly and for the rest is a
guess. It means that we assume that (\ref{defeta}) and (\ref{eta}),
with $\eta(s)$ calculated in sQED, still to some approximation accounts
for the effect. What we will actually do is to consider $\delta_{\rm
cut}^{\rm scan}(s)$, evaluated as just described, as a theoretical
uncertainty (model error).

Note that (\ref{scincl}) only contains the measured cross section
$\sigma^{\rm cut}_{\rm obs}(s)$, the known initial state correction
factors $\delta_{\rm ini}$ and $\rho_{\rm ini}^{\rm (cut)}$, and the
FSR--inclusive cross section $\sigma^{(\gamma)}_{\rm cut}(s)$,
which is the quantity to be extracted from the data since it
corresponds to the FSR--inclusive pion form factor via
(\ref{sigmaDressedCut}).

Whether the approximation $\hat{\sigma}^{(\gamma)}_{\rm cut}(s)$ obtained
via
\ba
\sigma_{\rm obs}^{\rm cut}(s) &=& \hat{\sigma}_{\rm cut}^{(\gamma)}(s)
\left[1+\delta_{\rm ini}(s,\Lambda)\right] \nn \\
&+&\int_{4m_\pi^2}^{s-2\sqrt{s}\Lambda} ds'\;
\hat{\sigma}_{\rm cut}^{(\gamma)}(s')\;\rho_{\rm ini}^{\rm cut}(s,s')
\; ,
\label{total3}
\ea
after neglecting $\delta_{\rm cut}^{\rm scan}(s)$ in
(\ref{scincl}), yields a good approximation
for the FSR--inclusive cross section is subject of the investigation
described in the following.

We first introduce the FSR--inclusive form factor
$F^{(\gamma)}_\pi(s)$ by
\be
|F^{(\gamma)}_\pi(s)|^2 = |F^{(0)}_\pi(s)|^2
\:\left(1+\frac{\alpha}{\pi}\eta(s)\right) +O(\alpha^2)\;,
\label{fpeta}
\ee
and {\em assume} that in some approximation it makes sense to write
formulas like (\ref{ff0}) also between $\sigma^{(\gamma)}(s)$ and
$F^{(\gamma)}_\pi(s)$ and between $\hat{\sigma}^{(\gamma)}(s)$ and
$\hat{F}^{(\gamma)}_\pi(s)$. Hard photon effects spoil these
assumptions at some level, but this at the moment is difficult to
quantify. So in the following, this will be part of our model
assumption (see below). 

To estimate the model dependence for the extraction of
$\sigma^{(\gamma)}_{\rm cut}(s)$ we
first generate a sample $\sigma_{\rm obs}^{\rm cut}(s)$, using the
pion form factor $|F_{\pi}^{(0)}(s)|^2$ as given
in~\cite{Akhmetshin:2001ig} and the relations (\ref{scincl}) --
(\ref{eta}) and (\ref{cut0}) -- (\ref{rhos}).
Then we utilize the MINUIT package~\cite{minuit} to obtain
$|\hat{F}_{\pi}^{(\gamma)}(s)|^2$ [which corresponds to
$\hat{\sigma}^{(\gamma)}_{\rm cut}(s)$]
from the $\sigma_{\rm obs}^{\rm cut}(s)$ data using (\ref{total3}).
For the data fitting we adopt again the Gounaris-Sakurai type parameterization
of $|F_{\pi}^{(0)}(s)|^2$ in the version proposed
in~\cite{Akhmetshin:2001ig}.

We then estimate the model error by
\ba
\Delta^{\rm scan}_{\rm cut}(s) &=&
\frac{\sigma^{(\gamma)}_{\rm cut}(s)-\hat{\sigma}^{(\gamma)}_{\rm cut}(s)}
{\sigma^{(\gamma)}_{\rm cut}(s)}
=\frac{|F_{\pi}^{(\gamma)}(s)|^2-|\hat{F}_{\pi}^{(\gamma)}(s)|^2}
{|F_{\pi}^{(\gamma)}(s)|^2}\;. \nn\\
\label{errorscanA}
\ea
\begin{center}
\begin{figure}[h]
\vspace*{-1.9cm}
\mbox{\epsfxsize 8.5cm \epsfysize 8.5cm \epsffile{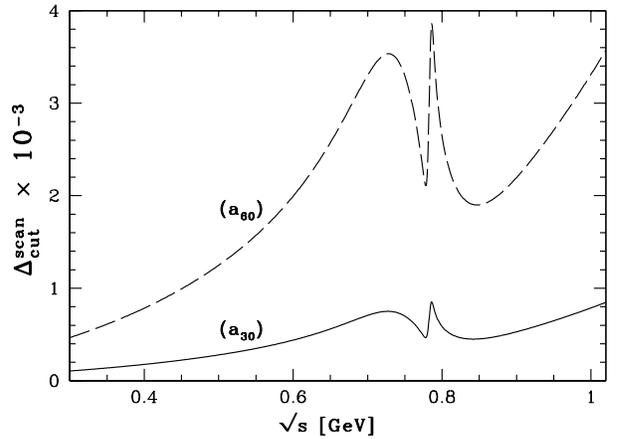}}
\caption{Estimated relative model error
for the extraction of the absolute square of the
FSR--inclusive pion form factor,
$|F_{\pi}^{(\gamma)}(s)|^2$, as a function of the center of mass energy
in a photon--inclusive scan experiments for different
C-symmetric angular cuts on the pion
angles. Curve
$(\rm a_{30})$ and curve $(\rm a_{60})$ corresponds to the cuts
$\Theta_\pi \geq 30^o$ and $\Theta_\pi \geq 60^o$, respectively.}
\label{scancut}
\end{figure}
\end{center}

\vspace*{-1cm}

\begin{center}
\begin{figure}[th]
\vspace*{-1.5cm}
\mbox{\epsfxsize 8.5cm \epsfysize 8.5cm \epsffile{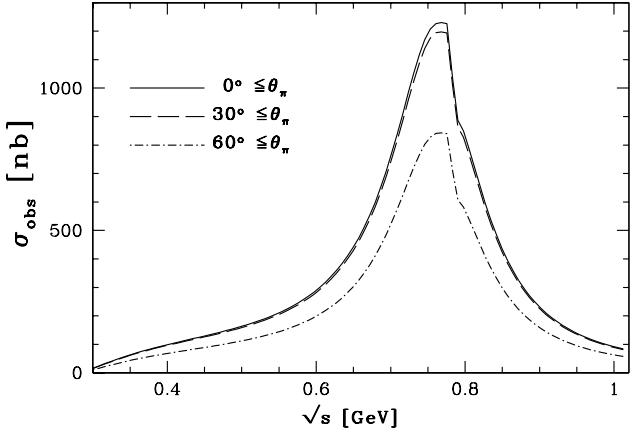}}
\caption{Observed total cross section for pion pair production as a
function of the center of mass energy ($\sqrt{s} \leq 1.02\;\GeV$) for the
considered cut scenarios.}
\label{crossscancut}
\end{figure}
\end{center}

The model error for the extraction of $|F_\pi^{(\gamma)}(s)|^2$ from
scan data with C-symmetric cuts is shown in Fig.~\ref{scancut}.
The detailed shape of the curve depends on the parameterization
of the pion form factor.
It can be noticed that the considered cuts do not lead to large model
errors. Even for the extreme cut of $\Theta_{\pi} \geq 60^o$
the model error is still below half a per cent, for $\Theta_{\pi} \geq 30^o$
it is below 1 per mill. In fact, the observed smallness of the model error is 
related to the p-wave-like angular distribution of the outgoing pions.
Fig.~\ref{crossscancut} shows that angular cuts of the pion angle
against the beam axis up to $30^o$ decrease the cross section
very little. 

\begin{center}
\begin{figure}[h]
\vspace*{-1.9cm}
\mbox{\epsfxsize 8.5cm \epsfysize 8.5cm \epsffile{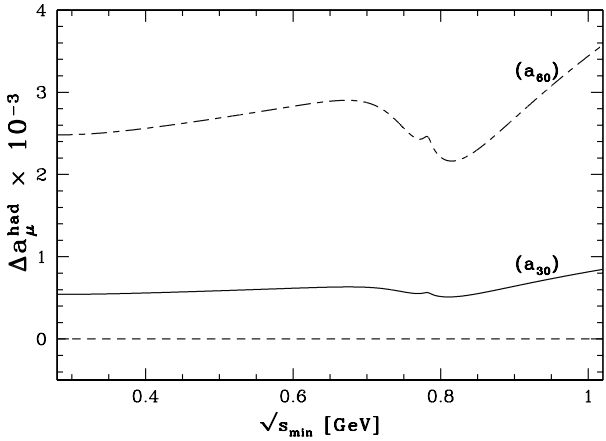}}
\caption{
Estimated relative model error in per mill
of $a_{\mu}^{\rm had}(s_{\rm min}<s<M_{\Phi}^2)$ in scan experiments.
The curves $(a_{30})$ and $(a_{60})$ correspond to the
cases in Fig.~\ref{scancut}. }
\label{amuerror}
\end{figure}
\end{center}

In Fig.~\ref{amuerror} the impact of the discussed model error
on $a_\mu^{\rm had}$
\ba
a_{\mu}^{\rm had} &=&
\left(\frac{\alpha m_\mu}{3\pi}\right)^2 \int_{4m_\pi^2}^{\infty}
ds \;\frac{R_{\pi\pi}(s)\hat{K}(s)}{s^2} \;,
\label{amudisp}
\ea
is shown.
For this we compare the values of $a_{\mu}^{\rm had}$ when inserting
\ba
R_{\pi\pi}(s) &\equiv&
R_{\pi\pi}^{(\gamma)}(s) = \frac{\beta_{\pi}^3}{4}\;|F_{\pi}^{(\gamma)}(s)|^2
\ea
%
\vspace*{-1cm}
into (\ref{amudisp}) [this value is denoted by
$a_{\mu}^{\rm had(\gamma)}$] with the value when inserting
\ba
R_{\pi\pi}(s) &\equiv&
\hat{R}_{\pi\pi}^{(\gamma)}(s) = \frac{\beta_{\pi}^3}{4}\;
|\hat{F}_{\pi}^{(\gamma)}(s)|^2 \;,
\ea
[this value is denoted by $\hat{a}_{\mu}^{\rm had(\gamma)}$].
Then the model error of $a_{\mu}^{\rm had}$ which is plotted
in Fig.~\ref{amuerror} for the curves $(a_{30})$ and ($a_{60}$) is defined as
\ba
\Delta a_{\mu}^{\rm had} = 
\frac{a_{\mu}^{\rm had(\gamma)}-\hat{a}_{\mu}^{\rm
had(\gamma)}}{a_{\mu}^{\rm had(\gamma)}}\; .
\label{amuerrorcurveA}
\ea
\vspace*{-1cm}

Let us remind that the present theoretical error is at the level of
1.2\% percent.

As a first summary we may say that the direct extraction of
$|\hat{F}_{\pi}^{(\gamma)}(s)|^2$ in an inclusive scan yields a very good
approximation of $|F_{\pi}^{(\gamma)}(s)|^2$, especially in the low
energy region where the contribution of FSR becomes large.
\subsection{Exclusive scenario}
Scan measurements in the past attempted to extract the ``bare'' cross
section $\sigma^{(0)}(s)$, undressed from photon radiation effects. 
As already mentioned, the bare cross section is the object of primary 
theoretical interest, in
principle. It is the quantity which allows us to extract the pion
form factor which encodes the strong interaction structure of the pion
in a world where the electromagnetic interaction has been switched
off. It is the non-perturbative quantity which one would
compute by a simulation in lattice QCD or investigate by means of
general low energy properties of the strong interactions like chiral
perturbation theory, locality and analyticity~\cite{Heiri}. The
theoretical concept of disentangling effects from different
interactions has been very successful, however, has its limitation at
some point. In phenomenology of low energy hadrons it is in fact not
possible to separate in a model-independent way QED from QCD effects
(at the level of accuracy we are considering here weak interaction
effects are negligible). This will be discussed in more detail in the
next section. Here, for the moment, we assume that
$\sigma^{(0)}(s)$ is a sensible pseudo observable.

\begin{center}
\begin{figure}[h]
\vspace*{-1.5cm}
\mbox{\epsfxsize 8.5cm \epsfysize 8.5cm \epsffile{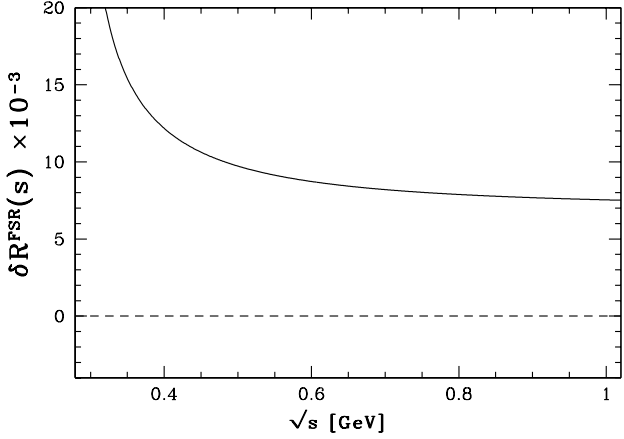}}
\caption{Importance of FSR in sQED.
The curve shows the difference between the absolute square
of the undressed and of the FSR--inclusive
pion form factor [see (\ref{errorscanB})].}
\label{scQEDa}
\end{figure}
\end{center}

Fig.~\ref{scQEDa} shows the relative deviation of the FSR--inclusive
pion form factor $|F_{\pi}^{(\gamma)}(s)|^2$ (being the
desired quantity to be inserted into the dispersion integrals)
from the undressed pion form factor $|F_{\pi}^{(0)}(s)|^2$, as calculated
within sQED [see (\ref{siggam}),(\ref{cut0}),(\ref{sigmaDressedCut})]
\ba
\delta R^{\rm FSR}(s) &=& \frac{\sigma^{(\gamma)}(s)-\sigma^{(0)}(s)}
{\sigma^{(0)}(s)} =
\frac{|F_{\pi}^{(\gamma)}(s)|^2-
|F_{\pi}^{(0)}(s)|^2}{|F_{\pi}^{(0)}(s)|^2} \nn\\
&=&\frac{\alpha}{\pi} \eta(s) + O(\alpha^2).
\label{errorscanB}
\ea
This quantity can be taken as a measure of the importance of FSR.

\vspace*{-1cm}

Since the analysis presented so far does not account for the fact that
generalized sQED describes correctly the soft photon part of the
FSR--spectrum only\footnote{Note that for $s\to 4 m_\pi^2$ real FSR is
known precisely since there is only enough phase space for the
radiation of soft photons and for soft photons the FSR radiation
mechanism is universal for given masses and charges of the final state
particles. Therefore estimating the model uncertainty to be given by
the sQED result appears to be too crude as it overestimates the
model uncertainty in the soft photon region.}, we would like to go
further and compare sQED modeling with a second one which
differs from it for hard photons.
As in~\cite{Hoefer:2001mx} we compare two
different treatments of final state corrections: once we take
for $\rho_{\rm fin}(s,s')$ and $\delta_{\rm fin}(s,\Lambda)$
the functions related to photonic radiation from point-like,
scalar pions and once we use the corresponding functions
for the photonic radiation from point-like, fermionic pions with
the same charge and mass (see Appendix~B).
The reason we do this is that in the soft photon limit the
scalar as well as the fermionic approach yields the same correct result
for the real photon contribution.
For hard photons on the other hand both scenarios are obviously
different. This will allow us to get some feeling
in what kinematic regions hard photons play a substantial role and
there lead to a large model uncertainty.

We thus consider in the following the model dependence of FSR effects,
by replacing the integrated final state corrections for scalar
particles, represented by the factor $\eta(s)$, by a corresponding
factor for a fermionic final state, which we denote by $\eta^f(s)$
(see again Appendix~B for explicit formulas). 
Though both $\eta(s)$ and $\eta^f(s)$
diverge when approaching the pion pair
production threshold (Coulomb pole\footnote{The Coulomb resummation 
has been considered in~\cite{Hoefer:2001mx}}), their difference in this limit
is a small number \newline $\lim_{s\to 4 m_\pi^2}[\eta(s)-\eta^f(s)]=1$ (in
units $\frac{\alpha}{\pi}$)\footnote{At high
energies scalar and fermionic FSR read
\ba
\eta(s\to\infty) &=& 3 \quad \rm and \quad
\eta^f(s\to\infty) = \frac{3}{4} \;. \nn
\ea
}.
As already mentioned, obviously, the generalized
sQED/fQED modeling of FSR obtained by replacing the
point--pion from--factor ``$1$'' by a form factor function
$|F_{\pi}(s)|^2$ of one single variable $s$ is valid for soft photons
only. A method which will allow us to describe also hard photons
in a realistic manner will be presented in a forthcoming paper.
\begin{center}
\begin{figure}[h]
\vspace*{-1.5cm}
\mbox{\epsfxsize 8.5cm \epsfysize 8.5cm \epsffile{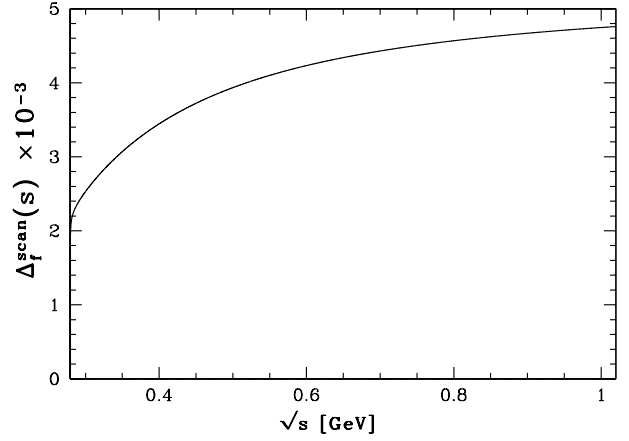}}
\caption{Model estimate of the relative model uncertainty
for the extraction of the absolute square of the
FSR--inclusive pion form factor,
$|F_{\pi}^{(\gamma)}(s)|^2$, as a function of the center of mass energy
in scan experiments.}
\label{fermSc}
\end{figure}
\end{center}

\vspace*{-1cm}

Our discussion thus motivates the consideration of the following measure
of the model dependence:
\ba
\Delta_f^{\rm scan}(s) &=& \frac{\sigma^{(\gamma)}-\sigma^{(\gamma),f}}
{\sigma^{(\gamma)}} =\frac{|F_{\pi}^{(\gamma)}(s)|^2-
|F_{\pi}^{(\gamma),f}(s)|^2}{|F_{\pi}^{(\gamma)}(s)|^2} \nn \\
& = &
\frac{\alpha}{\pi} [\eta(s)-\eta^f(s)] + O(\alpha^2).
\label{errorscanC}
\ea
It compares in a ratio $|F_{\pi}^{(\gamma)}(s)|^2$, being the pion
form factor dressed by scalar FSR, with the absolute square of the
pion form factor $|F_{\pi}^{(\gamma),f}(s)|^2$, being dressed by
fermionic FSR and corresponds to our ignorance of
FSR. Fig.~\ref{fermSc} shows $\Delta_f^{\rm scan}(s)$ as a function of
energy and suggests an uncertainty below 0.5\% over the whole energy
range of interest.

For the data analysis at the CMD-2 experiment \cite{Akhmetshin:2001ig}
the event selection was such that only events containing real low
energy photons were taken into account and thus the pions were
approximately back to back. Since for soft photons the FSR mechanism
is known (factorization), the real photonic corrections together with
the universal soft plus virtual IR terms can be subtracted from
the observed cross section in an essentially model independent
way\footnote{Collinear hard photons can be easily separated via the event
shapes.}. Accordingly, in order to keep the formulas simple, we define a
subtracted cross section $\sigma_{\rm obs}^{\rm subtr}(s)$ which does
not depend on the cuts any longer and is obtained from the
experimentally observed exclusive cross section in a theoretically
well controlled manner.

While real hard photons may be eliminated by appropriate cuts, the
same cannot be done for the remaining virtual corrections which
include high momentum scales in loops and hence their treatment is
model dependent. As a consequence the determination of the undressed
cross section $\sigma^{(0)}(s)$ as well as of the FSR--inclusive cross
section $\sigma^{(\gamma)}(s)$ from the given data suffers from model
dependence which is hard to estimate.

\begin{center}
\begin{figure}[h]
\vspace*{-1.5cm}
\mbox{\epsfxsize 8.5cm \epsfysize 8.5cm \epsffile{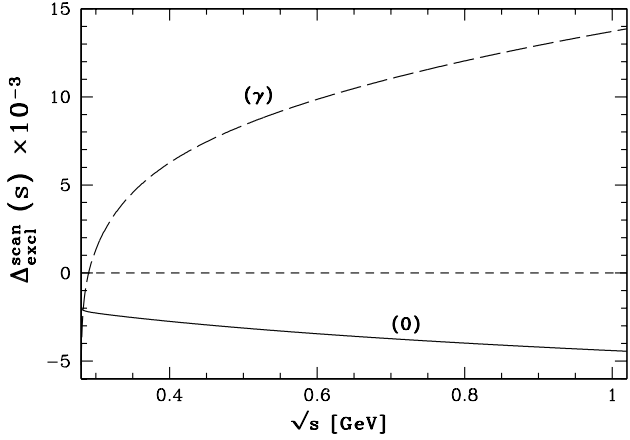}}
\caption{
Model error estimations for the extraction of the absolute square of the
pion form factor in an exclusive scenario
[see (\ref{delta0}), (\ref{deltag})]}
\label{cmd2fig}
\end{figure}
\end{center}

Here we will present a model error estimate assuming that the observed
cross section can be written as a product of $\sigma^{(0)}(s)$
containing all QCD effects (pion form factor) and a function
containing only the initial and final state QED corrections.  This ad
hoc assumption, although criticizable, seems to be the best we can do
so far.  Applying the procedure of real and IR photon
subtraction as described above then yields
\ba
\sigma_{\rm obs}^{\rm subtr}(s)
& \simeq &
\sigma^{(0)}(s)\;\left[1+\tilde{\delta}_{\rm ini}^{V+S}(s)+
\tilde{\delta}_{\rm fin}^{V+S}(s)\right]
 \label{gener}
\ea
which is the $O(\alpha)$ cross section including only the non--IR
initial and final state soft plus virtual corrections corresponding
to $\tilde{\delta}_{\rm ini}^{V+S}(s)$ and
$\tilde{\delta}_{\rm fin}^{V+S}(s)$.
Here of course the final state correction
factor $\tilde{\delta}_{\rm fin}^{V+S}(s)$ is not kn\-own.
Thus at this stage we have two unknowns: ${\sigma}^{(0)}(s)$ and the
FSR correction.
Only after assuming that FSR is given by sQED or fQED we can
then extract the undressed or the FSR--inclusive cross section
from $\sigma_{\rm obs}^{\rm subtr}(s)$
via the following formulas (see Appendix~B):
\ba
\hat{\sigma}^{(0)(,f)}(s) &=&
\sigma_{\rm obs}^{\rm subtr}(s)\;\frac{1}
{1+\tilde{\delta}_{\rm ini}^{V+S}(s)+\tilde{\delta}_{\rm fin(,f)}^{V+S}(s)},
\nn \\
&& \label{s0f}
\\
\hat{\sigma}^{(\gamma)(,f)}(s) &=&
\sigma_{\rm obs}^{\rm subtr}(s)\;\frac{1+\frac{\alpha}{\pi}\;
\eta^{(f)}(s)}{1+\tilde{\delta}_{\rm ini}^{V+S}(s)
+\tilde{\delta}_{\rm fin(,f)}^{V+S}(s)}. \nn\\
&& \label{setaf}
\ea
Using (\ref{s0f}) we could try to estimate the uncertainty
for the extraction of $\hat{\sigma}^{(0)}(s)$ via
\ba
\Delta_{\rm scan,0}^{\rm excl.}(s) &=&
\frac{\hat{\sigma}^{(0)}(s)-\hat{\sigma}^{(0),f}(s)}
{\hat{\sigma}^{(0)}(s)} \nonumber \\
&=& 1-\frac{1+\tilde{\delta}_{\rm ini}^{V+S}(s)
+\tilde{\delta}_{\rm fin}^{V+S}(s)}
{1+\tilde{\delta}_{\rm ini}^{V+S}(s)+\tilde{\delta}_{\rm fin}^{V+S,f}(s)} \;.
\label{delta0}
\ea
The such estimated model error is shown in Fig.~\ref{cmd2fig}
[curve (0)]. We would like to stress that we should be careful
not to take this error estimate obtained from the comparison of
two factorizable models too seriously.
In fact, if we would make the analogous comparison for the extraction
of $\hat{\sigma}^{(\gamma)}(s)$ the such obtained error
would be of the level of 1 per mill.
However, we would expect a larger error from non--factorizable FSR
contributions which cannot be estimated.

As an alternative possibility we may try, in the spirit of
(\ref{main}), what we get if we just correct for the
model--independent ISR and the model--independent IR--sensitive part of
FSR, obtaining
\ba
\tilde{\sigma}^{(\gamma)}(s) &=&
\frac{\sigma_{\rm obs}^{\rm subtr}(s)}{1+\tilde{\delta}_{\rm ini}^{V+S}(s)}.
\ea

Defining
\ba
\Delta_{\rm scan,\gamma}^{\rm excl.}(s) &=&
\frac{\tilde{\sigma}^{(\gamma)}(s)-
\hat{\sigma}^{(\gamma)}(s)}
{\tilde{\sigma}^{(\gamma)}(s)} \nonumber \\
&=& 1-\frac{[1+\frac{\alpha}{\pi}\eta(s)][1+\tilde{\delta}_{\rm ini}^{V+S}(s)]}
{1+\tilde{\delta}_{\rm ini}^{V+S}(s)+\tilde{\delta}_{\rm fin}^{V+S}(s)}
\label{deltag}
\ea
we can get a feeling for how well
$\tilde{\sigma}^{(\gamma)}(s)$
approximates the true ${\sigma}^{(\gamma)}(s)$ 
[estimated here by~(\ref{setaf})].
The result is shown in Fig.~\ref{cmd2fig} [curve ($\gamma$)].

To summarize: what can we get from a hard--photon exclusive
measurement:\\ {\bf i)} $\sigma^{(0)}$: in spite of the fact that all
real hard photons have been eliminated by cuts a surprisingly large
model uncertainty due to hard virtual photons poses an inherent
limitation: the corresponding uncertainty cannot fall below the level
of about 0.5\% (sQED). Strictly speaking $\sigma^{(0)}$ is not accessible
to experiment or only at limited precision by the fact that we cannot
switch off virtual QED effects in reality.

{\bf ii)} $\sigma^{(\gamma)}$: the missing real hard photons must be
calculated from a model like sQED and added by hand. What we get is a
model dependent $\hat{\sigma}^{(\gamma),({\rm model})}(s)$. Surprisingly,
the model dependence we estimate by our method (assuming factorization
with a single scale form factor) for this object is much smaller (at
the level of 0.1\% only). On the one hand this reduced model dependence
can be traced back to the Kinoshita-Lee-Nauenberg (KLN) theorem, which
infers that radiative corrections for total inclusive cross sections
are free from large logs. On the other hand it is not conceivable that
we get a more precise knowledge of $\sigma^{(\gamma)}$ from not
measuring hard photons than from actually measuring everything. In the
latter case the uncertainty shown in Fig.~\ref{scancut} has been estimated,
which, as expected, shows an increasing uncertainty for increasingly
strong cuts.  Nevertheless, even so we think that our method of
estimating the model dependence underestimates the error in the exclusive case,
it is a quantity which is protected by the KLN theorem from large
effects and thus is a quantity which seems to be under much better control
than e.g. the bare $\sigma^{(0)}$. How much better is hard to quantify
at this stage.

{\bf iii)} $\tilde{\sigma}^{(\gamma)}(s)$: is model independent per
definition but it is not the quantity of actual interest, as it is not
a good approximation to $\sigma^{(\gamma)}$. After all the hard real
photons are missing here and again we only can get what we are
interested in by adding the missing piece using a model. If we do so
we end up with $\hat{\sigma}^{(\gamma),({\rm model})}(s)$ again, up to
higher order terms. Then we are essentially back at ii).

\vspace*{-1cm}

\section{Model errors in radiative return measurements}
At radiative return experiments the spectral function
\newline $d\sigma/ds'$
is measured where $\sqrt{s'}$ is the invariant
mass of the non--photonic final state.
Let us first have a look at our ``master formula'' (\ref{main})
which is the photon--inclusive cross section in form of
a convolution integral in the
integration variable $s_V$, $s_V$ being
the invariant mass square of the hadronic final state including
the FSR photons but excluding the ISR photons.
One could think that, due to the factorization of ISR and FSR already
on the matrix element level [see (\ref{matrixelement})],
it is possible to extract $\sigma^{(\gamma)}(s_V)$ also from a
radiative return measurement at fixed $s$. Of course by rewriting
(\ref{main}) we can formally get
\ba
\frac{d\sigma_{\rm incl}}{ds_V} &=&
\sigma^{(\gamma)}(s_V)\;\rho_{\rm ini}^{\rm incl}(s,s_V)
+O(\alpha^2)_{\rm IFS}
.
\label{radret}
\ea
However, we cannot extract $\sigma^{(\gamma)}(s_V)$ from
the experimental data for the simple reason that $s_V$ is not an
observable. This can be immediately seen
already for the case of single photon emission where we have
$s_V=s'$
if the photon is emitted from the initial state, but $s_V=s$
if the photon is emitted from the final state.
Since on an event level ISR and FSR photons cannot be
distinguished, $s_V$ cannot be obtained from the observables
$s$ and $s'$.
The error we would make when identifying $s'$ with $s_V$ is
therefore of leading order FSR. This is one way to see that not only
there is no way to measure the FSR--inclusive cross section
$\sigma^{(\gamma)}(s)$ in a radiative return measurement but, in fact, we
have to deal with an $O(1)$ FSR--background leading to an in general
significant model error for the extraction of even the undressed cross section
$\sigma^{(0)}(s)$.
In the following we therefore are going to investigate the model error for
radiative return scenarios in a separate analysis.

Taking into account radiative corrections in the approximation where
only the leading single photon radiation from the final state is
included, the observed spectral function can be expressed as the sum
of an ISR and an FSR contribution since the IFS contribution drops
out:
\ba
\left(\frac{d\sigma}{ds'}\right)_{\rm obs} = \sigma^{(0)}(s')\;
\tilde\rho_{\rm ini}(s,s')+ \sigma^{(0)}(s)\;\tilde\rho_{\rm fin}(s,s'). \nn\\
\label{rr1}
\ea
Again we refer to Appendix~B and \cite{Hoefer:2001mx} for the
explicit expressions. Because we are interested to measure the
FSR--inclusive cross section it is tempting to extract the quantity
\ba
\tilde{\sigma}^{(\gamma)}(s') &=&
\frac{1}{\tilde\rho_{\rm ini}(s,s')}\;
\left(\frac{d\sigma}{ds'}\right)_{\rm obs}\;,
\label{rr3}
\ea
which is obtained by just subtracting the ISR part and dropping the
model-dependent last term of (\ref{rr1}).
However, since what we really want to get is $\sigma^{(\gamma)}(s)$
we have to rewrite (\ref{rr1}) in terms of this true $O(\alpha)$
FSR--inclusive cross section. We easily find
\ba
\sigma^{(\gamma)}(s') &=&
\frac{1}{\tilde\rho_{\rm ini}(s,s')}\;
\left(\frac{d\sigma}{ds'}\right)_{\rm obs}
+\delta^{\rm r.r.}(s,s'),
\label{rr2}
\ea
with
\ba
\delta^{\rm r.r.}(s,s') &=&
-\frac{\tilde\rho_{\rm fin}(s,s')}{\tilde\rho_{\rm ini}(s,s')}\;
\sigma^{(0)}(s)+\frac{\alpha}{\pi}\eta(s')\; \sigma^{(0)}(s') . \nn\\
\label{drr}
\ea
We can see that the first term of (\ref{drr}) is of $O(1)$ but the
second of $O(\alpha)$, such that no cancellation up to higher order
terms is possible! The first term may be considered as a correction
term only in the region where we have a large enhancement of
$\sigma^{(0)}(s')$ by the $\rho$ resonance, in particular, in
comparison to the reference cross section $\sigma^{(0)}(s)$ at
$s=M^2_\Phi$ (or $M_B^2$). In addition, the mass effects of photon
radiation by the pions vs. the ones from the electron--positron system
in fact lead to quite some suppression of $\tilde\rho_{\rm fin}(s,s')$
in comparison to $\tilde\rho_{\rm ini}(s,s')$, both of which are of
$O(\alpha)$. We conclude that, in radiative return experiments, a
direct model independent extraction of the FSR--inclusive
$\sigma^{(\gamma)}(s')$ is not possible at $O(\alpha)$ precision in
the naive way just considered.

Since we try here to discuss $O(\alpha)$ corrections to
$\sigma^{(0)}(s')$, which in the radiative return scenario is given
by
\ba
\sigma^{(0)}(s') =
\frac{1}{\tilde\rho_{\rm ini}(s,s')}\;
\left(\frac{d\sigma}{ds'}\right)_{\rm obs}
-\frac{\tilde\rho_{\rm fin}(s,s')}{\tilde\rho_{\rm ini}(s,s')}\;
\sigma^{(0)}(s)\;, \nn\\
\label{resolved}
\ea 
one obvious deficiency of our starting equation (\ref{rr1}) are the
missing higher order corrections. In the resolved form
(\ref{resolved}) we are at the $O(1)$ level only and thus we are not
able to seriously address the question of an FSR--inclusive
measurement. Obviously we loose one order in $\alpha$ in a radiative
return measurement. A discussion beyond leading order FSR would be
possible only if the complete next order version of (\ref{rr1}) would
be available. The need for going to higher orders also leads to more
problems with the treatment of hard photons radiated by the pions. Of
course the limitations are coming from the fact that virtual hard
photon emission by the pions cannot be switched off and our limited
theoretical knowledge of the higher order contributions have their
drawback for a precise determination of $\sigma^{(0)}(s')$ or
$\sigma^{(\gamma)}(s')$. The basic problems and limitations discussed
above for exclusive scan measurements also apply for the radiative
return method. Nevertheless, a discussion on the basis of (\ref{rr1})
addresses the major difficulty we encounter in the attempt to measure the
FSR-dressed cross--section in a radiative return experiment.

As (\ref{rr2}) tells us, the FSR correction
$\frac{\alpha}{\pi}\eta(s')$, given precisely by the second term of
(\ref{drr}), is {\em completely lost} once we drop $\delta^{\rm
r.r.}(s,s')$ in order to get the model-independent quantity
(\ref{rr3}), which means that the latter is not a very meaningful
quantity. Therefore, in~\cite{Hoefer:2001mx} we proposed to unfold the
raw data from all photon radiation, by modeling FSR by generalized
sQED. The extracted undressed cross section $\sigma^{(0)}(s)$ then
suffers from model dependence.
From the preceding discussion we know what the actual problem is: in
first place we have to control the first term (\ref{drr}) which is
suppressed by \newline $\tilde\rho_{\rm fin}(s,s')/\tilde\rho_{\rm
ini}(s,s')$ and in regions where $\sigma^{(0)}(s)/\sigma^{(0)}(s')$ is
small but otherwise is of $O(1)$.

A measure for the relative importance of the disturbing
model--dependent FSR term $\delta^{r.r.}$ is
\ba
\delta r^{\rm FSR}(s',s) &=&
\frac{\tilde{\sigma}^{(\gamma)}(s')-\sigma^{(0)} (s')}
{\sigma^{(0)} (s')} \;,
\label{deltarr0}
\ea
which is indeed a measure for the importance of FSR as given by
the radiator function $\tilde\rho_{\rm fin}(s,s')$. It has to be
compared with (\ref{errorscanB}) which measures the FSR in the
integrated form (\ref{eta})
for the case of a scan experiment.
Obviously, (\ref{deltarr0}) definitely does not account
for the $\frac{\alpha}{\pi}\eta(s')$ term in
$\sigma^{(\gamma)}(s')$. We may consider it, however, as
a measure for the model--dependence of the extraction of the undressed
$\sigma^{(0)}(s')$. Thus let us point out once more, it would be
misleading to think that $\tilde{\sigma}^{(\gamma)}(s')$ in
any sense would approximate $\sigma^{(\gamma)}(s')$ to
better than the $O(1)$ level. Although it includes FSR effects, it does
not include the ones we are looking for.
\begin{center}
\begin{figure}[t]
\vspace*{-1.5cm}
\mbox{\epsfxsize 8.5cm \epsfysize 8.5cm \epsffile{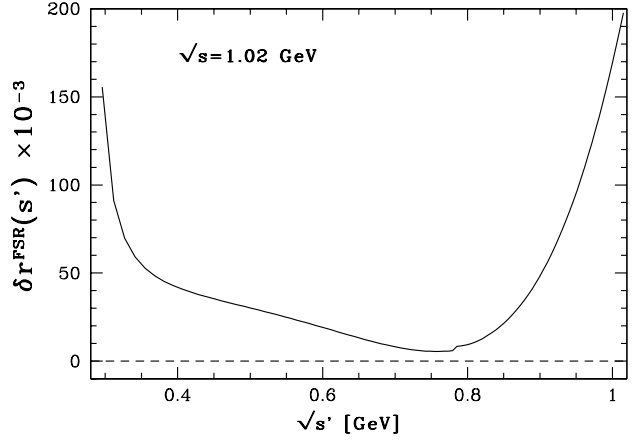}}
\caption{
Importance of FSR for sQED. The curve shows the scenario for the
extraction of $|F_{\pi}^{(0)}(s')|^2$ in radiative return experiments
($\sqrt{s}=M_{\Phi}$) estimated by once including and once excluding
final state corrections [See (\ref{deltarr0})].}
\label{scRR}
\end{figure}
\end{center}

\vspace*{-1cm}

The size of FSR effects for the radiative return
scenario is depicted in Fig.~\ref{scRR}.

Clearly the effect is large both in the soft ($s'\lapprox s$) and the
hard ($s'\ll s$) photon limits. It is small only where it is suppressed
by a large $\sigma^{(0)}(s')$, i.e., around the $\rho$--resonance.
\newline
In fact
the raise of $\delta r^{\rm FSR}(s',s)$ in the soft photon regime (for
$s' \to s$) just means that soft photon FSR effects become large and
does not imply a large model dependence. Hence taking 
(\ref{deltarr0}) as an estimation of the model error 
would be too rough. It does not yet take into account that sQED
describes well the soft photon regime. In order to get a more
realistic measure for the model dependence we have to proceed as in
the previous section.

Given the experimental distribution
$\left(\frac{d\sigma}{ds'}\right)_{\rm obs}$ in (\ref{rr1}) the
extracted $\sigma^{(0)}(s')$ depends on the unknown FSR radiator
$\tilde\rho_{\rm fin}(s,s')$, which we again model by sQED.
Analogously to our analysis for the scan scenario
[see (\ref{errorscanC})] we compare the
result for a scalar vs. a fermionic radiator by looking at
\ba
\Delta_f^{\rm r.r.}(s') &=& \frac{\sigma^{(0),f}(s')-\sigma^{(0)}(s')}
{\sigma^{(0)}(s')} \;,
\label{errorRRC}
\ea
with $\sigma^{(0)}(s')$ extracted assuming sQED and
$\sigma^{(0),f}(s')$
extracted assuming a fermionic radiator $\tilde\rho_{\rm fin}^{f}(s,s')$
[see (\ref{rhofinscal}) and (\ref{rhofinferm})].

\begin{center}
\begin{figure}[h]
\vspace*{-1.5cm}
\mbox{\epsfxsize 8.5cm \epsfysize 8.5cm \epsffile{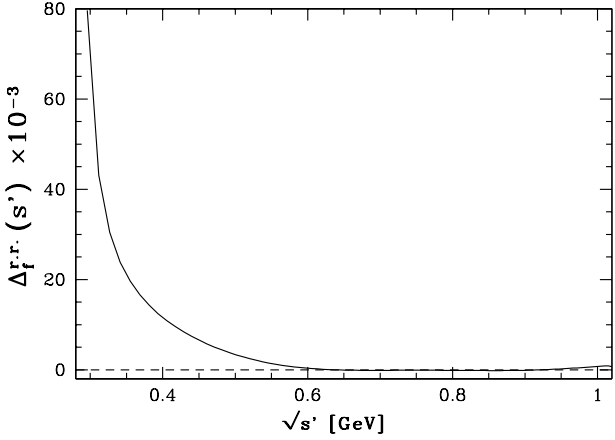}}
\caption{
Estimated relative model error for the extraction of
$|F_{\pi}^{(0)}(s')|^2$ in radiative return experiments
($\sqrt{s}=M_{\Phi}$). The curve describes a scenario
where the model error is estimated by a comparison of FSR once
from scalar particles and once from fermions of the same charge and mass.}
\label{fermRR}
\end{figure}
\end{center}

The result is shown in Fig.~\ref{fermRR}. We would like to stress once
more that (\ref{rr1}) does not incorporate the ISR$\otimes$FSR and IFS
effects, which account for an additional $O(\alpha)$ model error
contribution. Thus, Fig.~\ref{fermRR} when taken without this proviso
is misleading at energies there the $O(1)$ term is kinematically
suppressed to be smaller than the missing, presently unknown,
$O(\alpha)$ terms\footnote{Their evaluation would require a full
two--loop calculation of the process $\epmppm$.}, which are expected to
be at the few per mill level.

\vspace*{-1cm}

Clearly, the ``scalar versus fermionic'' scenario gives, as expected,
a small model dependence for the soft photon region but the model
error remains large for pion pair production near threshold energies
where hard photons are involved necessarily.

We repeat that for radiative return experiments we cannot see a
possibility to measure the FSR--inclusive cross section
$\sigma^{(\gamma)}(s)$, at least not in some obvious way. We only are
able to extract the undressed cross--section $\sigma^{(0)}(s)$. At
$O(\alpha)$ precision even this is only possible in a model dependent
way, since the problems are the same as the ones we have addressed
earlier for the exclusive scan measurements. To get
$\sigma^{(\gamma)}(s)$ we must add up the FSR as given by sQED, with
the drawback that we have to live with the model dependence as
illustrated by Fig.~\ref{fermRR}. Since the disturbing second term in
(\ref{rr1}) is much smaller for a radiative return experiment at a
$B$--factory, it seems that the chances to get a model--independent
determination of $\sigma^{(0)}(s')$ there, could be good for what
concerns the theoretical uncertainties associated with FSR.  It can be
easily checked that for such measurements at a $B$--factory indeed the
model error is limited by higher order FSR effects (not considered
here) since the $O(1)$ FSR contribution is essentially 0 due to the
fact that $\sigma_{\pi\pi}(s=M_{\Upsilon_{4S}}^2)/
\sigma_{\pi\pi}(s\leq 1\;\GeV) \lapprox 0.04$.
The observed cross section, on the other hand, is reduced by about two
orders of magnitude in respect to a $\Phi$--factory measurements.
However, for the BABAR experiment at SLAC this drawback is compensated
by a very high luminosity which is about a factor of $400$ larger than
at the $DA\Phi NE$ collider.  It is important to note that a good
control of ISR will be required since the gap between $s'$ and $s$ is
much larger than for $\Phi$--factories.  In particular the singlet
initial state pair production channel ``$e^+ e^- \to \pi^+ \pi^- e^+
e^-$'' yields the dominant contribution to the inclusive channel
``$e^+ e^- \to \pi^+ \pi^- \;+\; \rm anything$'' at $B$--factory
energies, being about a factor of 3 larger around the $\rho$ peak and
even a factor of about $30$ larger near $\pi\pi$ threshold than the
contribution from ``$e^+ e^- \to \pi^+ \pi^- \;+\; \rm photons$''.
Higher order photonic corrections also have to be taken into account.
Leading $\log$ photonic $O(\alpha^3)$ corrections here contribute
about $0.3$ per cent to the inclusive spectral function.

Let us now apply C-symmetric angular cuts to the calculations concerning
the radiative return method. Cut objects have been
defined in (\ref{cut0}) -- (\ref{rhos}).

Analogously to the case without cuts in (\ref{deltarr0}) the importance of
the FSR contribution for a C-symmetric
cut scenario can be defined as the relative deviation of
\ba
\tilde{\sigma}_{\rm cut}^{(\gamma)}(s') &=&
\frac{1}{\rho_{\rm ini}^{\rm cut}(s',s)}
\;\left(\frac{d\sigma}{ds'}\right)_{\rm obs,cut}
\ea
from the undressed cross section $\sigma_{\rm cut}^{(0)}$:
\be
\delta r^{\rm FSR}_{\rm cut}(s') =
\frac{\tilde{\sigma}^{(\gamma)}_{\rm cut}(s')
-\sigma_{\rm cut}^{(0)}(s')}{\sigma_{\rm cut}^{(0)}(s')} \;.
\label{deltarr3}
\ee
\begin{center}
\begin{figure}[h]
\vspace*{-1.5cm}
\mbox{\epsfxsize 8.5cm \epsfysize 8.5cm \epsffile{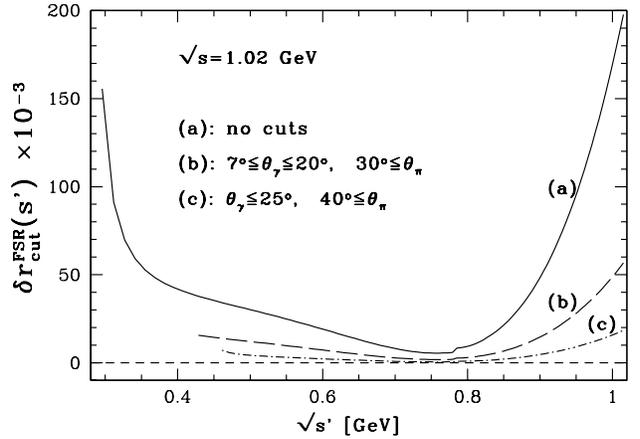}}
\caption{Relative FSR contribution as given by sQED obscuring
the extraction of $|F_{\pi}^{(0)}(s')|^2$ in radiative return
experiments ($\sqrt{s}=M_{\Phi}$) for different angular cuts
[see (\ref{deltarr3})]. Curve
(a) is the same as the curve in Fig.~\ref{scRR}. Curve (b)
corresponds to the cut scenario where only events are taken into
account for which the laboratory angle between the pion momenta and
the beam axis $\theta_\pi$ is larger than $30^o$ and the laboratory
photon angle $\theta_\gamma$ is restricted to a region
$7^o\leq\theta_\gamma\leq 20^o$. In a similar way curve (c)
corresponds to $\theta_\pi\geq 40^o$ and $\theta_\gamma\leq 25^o$.}
\label{naive}
\end{figure}
\end{center}


\begin{center}
\begin{figure}[h]

\vspace*{-1.5cm}

\mbox{\epsfxsize 8.5cm \epsfysize 8.5cm \epsffile{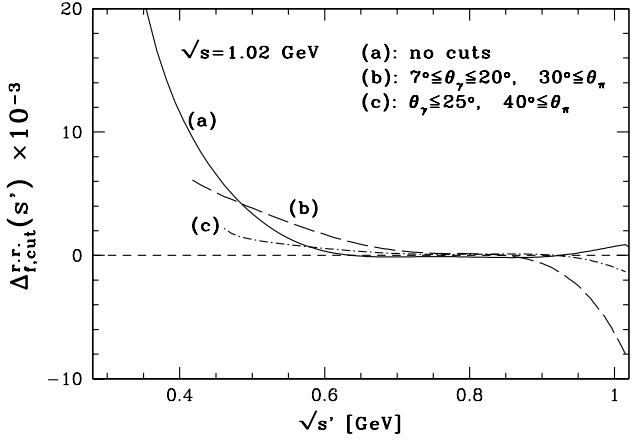}}
\caption{Estimated relative model error in per mill for the extraction of
$|F_{\pi}^{(0)}(s')|^2$
in radiative return experiments ($\sqrt{s}=M_{\Phi}$)
for different angular cuts. Here the scenario is shown where
the model error is estimated by a comparison of FSR from scalar particles
with FSR from fermions of the same charge and mass
[see (\ref{deltarrferm})]. The cut scenarios (a),
(b) and (c) are the same as in Fig.~\ref{naive}.}
\label{genius}
\end{figure}
\end{center}

\vspace*{-1.8cm}

\begin{center}
\begin{figure}[h]
\vspace*{-1.5cm}
\mbox{\epsfxsize 8.5cm \epsfysize 8.5cm \epsffile{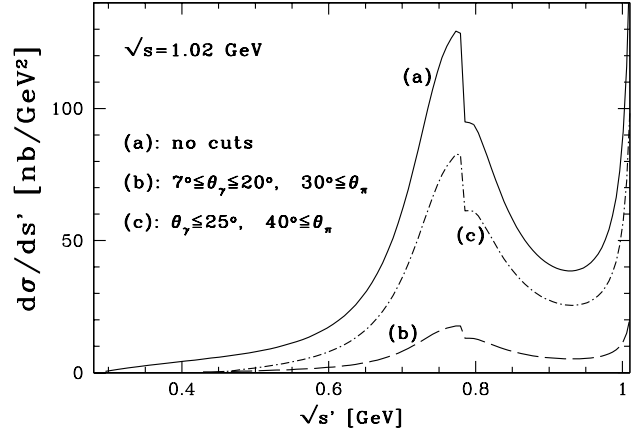}}
\caption{Pion pair invariant mass distribution in radiative return
experiments ($\sqrt{s}=M_{\Phi}$) for the three discussed cut scenarios.}
\label{cross}
\end{figure}
\end{center}

Again we may ``guesstimate'' a model uncertainty by replacing and
comparing the sQED model with the fQED model. In the
latter case we replace $\rho_{\rm fin}^{\rm cut}(s,s')$ by
\ba
\rho_{\rm fin,f}^{\rm cut}(s,s') &=& \frac{1}{
\sigma_{\rm 0,cut}^{\rm point,f}(s)}
\;\left(\frac{d\sigma}{ds'}\right)_{\rm fin,cut}^{\rm point,f} \;
\ea
where
\ba
\sigma_{\rm 0,cut}^{\rm point,f}(s) &=&
\pi\;\frac{\alpha^2\beta_{\pi}^3\:\cos\Theta_{\pi}^{\rm M}}{s}\; \left( {\frac{s+4 m_\pi^2}{s-4 m_\pi^2}
+\frac13 \cos^2{\Theta_\pi^{\rm M}}}\right) \; ,
\nn \\
\ea
$\Theta_{\pi}^{\rm M}$ being the minimal angle between the pion momenta and 
the beam axis allowed by the given cuts. 
In analogy to the case without cuts in (\ref{errorRRC}) we then may
define a model error as
\ba
\Delta^{\rm r.r.}_{\rm f,cut}(s') &=&
\frac{\sigma^{(0),f}_{\rm cut}(s')
-\sigma_{\rm cut}^{(0)}(s')}{\sigma_{\rm cut}^{(0)}(s')} \;.
\label{deltarrferm}
\ea
Fig.~\ref{naive} shows the FSR contribution and Fig.~\ref{genius} the
estimated model error for different kinematic cuts. It is
interesting to note that the FSR contribution can be clearly reduced
by the chosen kinematic cuts. This is especially obvious in the soft
photon region. The suppression of the model error, being estimated by
the scalar vs. fermionic scenario, by the considered cuts,
however, is not obvious. As a matter of fact the cross sections drop
out very quickly for low $\sqrt{s'}$ and vanish (up to higher order
effects) below $\sqrt{s'}\simeq
0.5$ GeV (see Fig.~\ref{cross}). 

\vspace*{-1cm}

Finally we would like to comment on the estimation of FSR
from a measurement of the pion forward-backward
asymmetry~\cite{karlsruhe},
defined as
\ba
\rm A_{FB}(s')
&=& \frac{\left(\frac{d\sigma}{ds'}\right)^{\cos\theta_\pi<0}
-\left(\frac{d\sigma}{ds'}\right)^{\cos\theta_\pi>0}}
{\left(\frac{d\sigma}{ds'}\right)^{\cos\theta_\pi<0}
+\left(\frac{d\sigma}{ds'}\right)^{\cos\theta_\pi>0}} \nn \\
&=&
\frac{\left(\frac{d\sigma}{ds'}\right)^{\cos\theta_\pi<0}_{\rm int}}
{\left(\frac{d\sigma}{ds'}\right)^{\cos\theta_\pi<0}_{\rm ini}
+\left(\frac{d\sigma}{ds'}\right)^{\cos\theta_\pi<0}_{\rm fin}} \;.
\label{eqAFB}
\ea
\begin{center}
\begin{figure}[h]
\vspace*{-1.5cm}
\mbox{\epsfxsize 8.5cm \epsfysize 8.5cm \epsffile{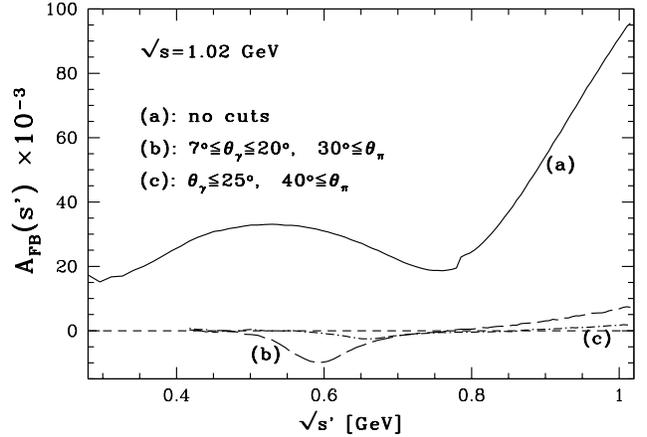}}
\caption{Forward-backward asymmetry in per mill of negatively charged pions in
radiative return experiments. $\theta_\pi$ is here the angle between the
momentum of the incoming $e^-$ and the outgoing $\pi^-$.}
\label{AFB}
\end{figure}
\end{center}

\vspace*{-1.4cm}

Here $\theta_\pi$ is the angle between the momentum of the incoming
$e^-$ and the outgoing $\pi^-$. One could think that if $A_{FB}$ is
small also the FSR contribution must be suppressed,
because, as can be seen from (\ref{eqAFB}), the IFS
contribution to the spectral function
$(d\sigma/ds')_{\rm int}$ determines the size of $A_{FB}$. The
smallness of $A_{FB}$, enhanced by kinematic cuts, could then be taken
as a measure for the suppression of FSR in experiments. In
Fig.~\ref{AFB} $A_{FB}$ is shown for the three different cut
scenarios. If we compare Fig.~\ref{AFB} with
Figs.~\ref{scRR}, \ref{fermRR}, \ref{naive} and \ref{genius} we observe
that we would get a different estimation of model dependence from
$A_{FB}$ than from our previous analysis of FSR. For example, for the
case of no cuts we get an increasing model dependence from our
investigation of FSR if $s'$ approaches threshold while $A_{FB}$
decreases for $s'\to 4m_\pi^2$.

Obviously, $A_{FB}$ is affected by FSR and hence can only be predicted
by assuming a model like sQED. A comparison of such a prediction with
experimental data is able to shed light on the validity of such a model.
E.g., if we observe a good agreement of the measured $A_{FB}$ with the
sQED prediction we could expect that likely also the FSR prediction 
by sQED could be a good approximation.

\begin{center}
\begin{figure}[h]
\vspace*{-1.5cm}
\mbox{\epsfxsize 8.5cm \epsfysize 8.5cm \epsffile{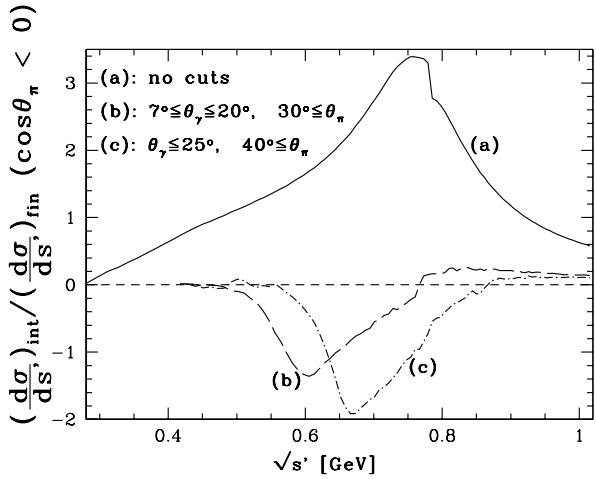}}
\caption{Ratio of the IFS contribution to final
state contribution for $\cos\theta_\pi<0$. Here
$\theta_\pi$ is the angle between the momentum of
the incoming $e^-$ and the outgoing $\pi^-$.}
\label{intfin}
\end{figure}
\end{center}

Of course $A_{FB}$ is not a direct measure
of FSR and hence of the model dependence related to it.
In fact the ratio of the FSR contribution and the IFS contribution
is a strongly varying function of $s'$ (see Fig.~\ref{intfin}).
While in the region of the $\rho$ resonance this ratio is relatively
large it becomes small for low $s'$. This is true no matter whether cuts are
applied or not.

\vspace*{-1cm}

\section{Conclusions}
The importance of FSR for the extraction of the pion form factor from
experimental data and for determining $a_\mu^{\rm had}$ has been discussed
both for scan and radiative return experiments.

We have shown that, by a photon--inclusive measurement and just subtracting
ISR, a direct extraction of the FSR--inclusive cross section 
$\sigma^{(\gamma)}$ in scan
experiments is possible with excellent accuracy.
In this case the model error is due to the breaking of ISR$\otimes$FSR
factorization by kinematic cuts and is of the order of a few per mill
for the discussed C-symmetric angular cuts.

On the other hand for exclusive measurements it is much more difficult to give
a reliable estimation of the model error. The main reason is that
without relying on ad hoc models like sQED
we are not able to disentangle the to be extracted QCD quantity
(pion form factor) from the real and virtual QED corrections.

For radiative return measurements a direct extraction of
$\sigma^{(\gamma)}(s)$ is not possible.  In fact we can neither obtain
$\sigma^{(\gamma)}(s)$ nor $\sigma^{(0)}(s)$ at $O(\alpha)$ FSR
precision without resorting to a model. Furthermore we have to deal
with an $O(1)$ FSR background which is under control only if one of
the following criteria apply: i) it can be subtracted by using
factorization in the soft photon region ($s'\simeq s$), ii) it is
suppressed by kinematic cuts (where it is possible) or iii) it is
negligible in regions where $\sigma^{(0)}(s') \gg \sigma^{(0)}(s)$.
We have shown that at $\phi$ factories like \dafne we can control this
FSR background only above $\sqrt{s'}\simeq 500$ MeV.

We also had a look at the pion forward--backward asymmetry which is a
model dependent quantity and thus is able to test model predictions
against reality. Suppose that data would agree well with the sQED
prediction this could be an indication that sQED also is able to
describe FSR to some extent. However, it is not possible to estimate
the FSR contribution in a straightforward way from a measurement of
$A_{FB}$. The ratio of the FSR and the IFS correction is a strongly
varying function of $\sqrt{s'}$. $A_{FB}$ becomes small for low
$\sqrt{s'}$ although the final state correction and hence the related
model dependence is large.

For radiative return measurements of $|F_\pi^{(0)}(s)|^2$ at
$B$--factories the $O(1)$ FSR background term is practically absent.
The model uncertainty due to FSR for the extraction of the undressed
pion form factor in this case is determined by higher order
effects. At BABAR the smallness of the observed cross section is
compensated by a high luminosity. Initial state corrections here have
to be known to a high precision. In particular for $\sqrt{s'} \leq 1$
GeV the dominant pion pair production channel is $e^+e^- \to
\pi^+\pi^- e^+e^-$.

In conclusion, to measure the FSR--inclusive pion form factor
$|F_{\pi}^{(\gamma)}(s)|^2$ precisely and in a model independent
manner at low energies, precise data from photon--inclusive scan
experiments will be indispensable.  We hope that such an experiment
will be possible at \dafne at a later stage.

\section*{Acknowledgments}
It is our pleasure to thank G. Colangelo, A.~Denig, S. Di Falco,
J. Gasser, W.~Kluge, A.~Nyffeler and G.~Venanzoni for
fruitful discussions. J.G.~was supported by the Polish Committee for
Scientific Research under Grant No. 2P03B05418.

\appendix

\section{Factorization of ISR and FSR}
\setcounter{equation}{0}

Consider the process ``$e^+e^-\to\gamma^*\to X + \rm photons$'' where X is an
arbitrary non--photonic final state and the photons can either be emitted
from the initial state (IS) or the final state (FS).
We are interested in an expression for the inclusive total
cross section $\sigma^{\rm incl}\equiv\sigma(e^+e^-\to X + \rm photons)$
in terms of the FSR--inclusive
 cross section that is dressed by all
real and virtual FS photonic corrections
$\sigma^{(\gamma)}$
and a universal IS
radiator function $\rho_{\rm ini}^{\rm incl}$
corresponding to all IS real and virtual corrections.
As will be shown in the following such a factorization is in fact
possible up to $O(\alpha^2)$ IFS real and virtual QED corrections.

Let us consider first the process $e^+e^-\to\gamma^*\to X + r\;\gamma$,
where $r$ is a given number of real photons which can be emitted either
from the IS or the FS.
The amplitude ${\cal M}^{(r)}$ corresponding to this process can
be written as the sum of all sub-amplitudes
${\cal M}^{(v_i,v_f)}_{(r_i,r_f)}$ corresponding to $r_i$ real and $v_i$
virtual photons attached to the IS $e^+e^-$ pair, $r_f$ real and $v_f$
virtual photons attached to the final state $X$,
and $v_{int}$ additional virtual
photons connecting the IS and the FS. For given $r$ we have the condition
$r_i+r_f=r$. In the following we will neglect box-like diagrams, thus
put $v_{int}=0$. Hence we only keep the pure IS and FS virtual
corrections (we will come back to the IFS contributions later).
The IR divergences of the virtual corrections are assumed
to be regularized by a small photon mass.
Without such an IR-regulator ${\cal M}^{(r)}$ would not be defined.
Obviously ${\cal M}^{(r)}$ by itself does not correspond
to a physical observable. However, at the end, after summation of all
the contributions corresponding to all
sub--amplitudes, we will obtain IR--finite, physical quantities and
the IR regulator can be removed.
The amplitude corresponding to the emission of $r$ real photons
can now be written as
\ba
{\cal M}^{(r)} &=& \sum_{r_{i,f}}\sum_{v_{i,f}=0}^\infty
\left. {\cal M}^{(v_i,v_f)}_{(r_i,r_f)}\right|_{r_i+r_f=r}
\nn\\
&=&
\sum_{r_{i,f}}\sum_{v_{i,f}=0}^\infty
\left[ A_{\mu}^{(r_i,v_i)}(q,q_V,\{k^{(i)}\}) \right. \nn\\
&\times & \left.
\;B^\mu_{(r_f,v_f)}(q_V,\{k^{(X)}\},\{k^{(f)}\})\right]_{r_i+r_f=r} ,
\label{amplitude}
\ea
where $\{k^{(i)}\}=\{k^{(i)}_1\dots k^{(i)}_{r_i}\}$
are the IS real photon momenta,
$\{k^{(f)}\}=\{k^{(f)}_1\dots k^{(f)}_{r_f}\}$
the FS real photon momenta,
$\{k^{(X)}\}=\{k^{(X)}_1\dots k^{(X)}_{n_X}\}$
the momenta of the non-photonic FS particles,
$q=p_1+p_2$ the sum of the incoming $e^+e^-$ momenta
and finally $q_V = q - \sum k^{(i)}=\sum k^{(X)}+\sum k^{(f)}$
the momentum of the
virtual photon $\gamma^*(q_V)$ connecting the IS and the FS.
In Eq.~(\ref{amplitude}) we have written the sub-amplitudes
${\cal M}^{(v_i,v_f)}_{(r_i,r_f)}$ as contractions of
rank-1 tensors $A_{\mu}^{(r_i,v_i)}$ containing the IS real and virtual
corrections with the rank-1 tensors $B^\mu_{(r_f,v_f)}$ containing
the FS real and virtual corrections and the photon propagator
related to $\gamma^*(q_V)$. This factorization of the amplitude
is only possible because we have neglected the virtual IFS contributions.
Defining
\ba
&&\tilde{A}_{\mu}^{(r_i)}(q,q_V,\{k^{(i)}\})
= \sum_{v_i=0}^\infty A_{\mu}^{(r_i,v_i)}(q,q_V,\{k^{(i)}\}) \;, \nn\\
&&\tilde{B}^\mu_{(r_f)}(q_V,\{k^{(X)}\},\{k^{(f)}\})
= \nn \\
&& \qquad\qquad\quad
\sum_{v_f=0}^\infty B^\mu_{(r_f,v_f)}(q_V,\{k^{(X)}\},\{k^{(f)}\}) \;,
\ea
thus summing over all virtual IS and FS virtual corrections,
we can write ${\cal M}^{(r)}$ simply as
\ba
{\cal M}^{(r)} &=& \left\{\biggl[\sum_{r_{i}}
\tilde{A}^{(r_i)}_{\mu}(q,q_V,{k^{(i)}})
\biggr] \right. \nn\\
&\times & \left. \biggl[\sum_{r_{f}}\tilde{B}^\mu_{(r_f)}
(q_V,{k^{(X)}},{k^{(f)}})\biggr] \right\}_{r_i+r_f=r} \;.
\ea
Let us remember that ${\cal M}^{(r)}$ contains all IS and FS virtual
corrections
to the given channel with $r$ real (IS or FS) photons.
Squaring the amplitude, averaging over the incoming $e^+e^-$ spins and
summing over the spins of the ISR photons $s_{(i)}$, the spins of
the FSR photons $s_{(f)}$ and the spins of the non-photonic particles
$s_{(X)}$ yields
\ba
\ol{|{\cal M}^{(r)}|^2} &=& \frac{1}{4} \sum_{s_{(X)}}
\sum_{s_{(i)}}\sum_{s_{(f)}}
\sum_{r_i,r_f} |{\cal M}^{(r_i,r_f)}|^2 \quad + \quad \rm IFS\;terms \nn\\
&=&
\sum_{r_i,r_f} \ol{|{\cal M}^{(r_i,r_f)}|^2} \quad + \quad \rm IFS\;terms \;,
\ea
where ``IFS terms'' corresponds to real photon IFS contributions.
Since we neglected already the virtual IFS contributions we have to do
the same for the real IFS contributions. Otherwise we would have
no cancellation of the IR divergences if later, after summing up all
contributions $r=0\dots\infty$, we want to
remove the IR regulator.
Neglecting the IFS contributions
we can now express the squared amplitude $\ol{|{\cal M}^{(r)}|^2}$
as a sum of the incoherent contributions $\ol{|{\cal M}^{(r_i,r_f)}|^2}$
which can be written as a contraction of an IS and a FS tensor,
respectively:
\ba
\ol{|{\cal M}^{(r_i,r_f)}|^2} &=& E_{\mu\nu}^{(r_i)}(q,q_V,\{k^{(i)}\})
F^{\mu\nu}_{(r_f)}(q_V,\{k^{(X)}\},\{k^{(f)}\}) \;, \nn\\
\label{matrixelement}
\ea
with
\ba
&&E_{\mu\nu}^{(r_i)}(q,q_V,\{k^{(i)}\}) = \\
&& \;\;\;\;\;\;\;\;\;\; \frac{1}{4} \sum_{s_{(i)}}
\tilde{A}_{\mu}^{(r_i)}(q,q_V,\{k^{(i)}\})
\;\tilde{A}_{\nu}^{(r_i)*}(q,q_V,\{k^{(i)}\})
\;,\nn \\
&&F^{\mu\nu}_{(r_f)}(q_V,\{k^{(X)}\},\{k^{(f)}\}) = \\
&&\;\;\;\;\;\;\;\;\;\; \sum_{s_{(f)}}\sum_{s_{(X)}}
\tilde{B}^\mu_{(r_f)}(q_V,\{k^{(X)}\},\{k^{(f)}\}) \nn\\
&& \;\;\;\;\;\;\;\;\;\;\; \;\;\;\;\;\;\;\;\;\;\; \times
\tilde{B}^{*\nu}_{(r_f)}(q_V,\{k^{(X)}\},\{k^{(f)}\}) \;. \nn
\ea
Note that $E_{\mu\nu}^{(r_i)}(q,q_V,\{k^{(i)}\})$ now only contains
the IS real and virtual corrections while $F^{\mu\nu}_{(r_f)}$
only contains the FS real and virtual corrections.
Thus, already at this stage we obtain complete factorization of the IS
and the FS corrections.
Phase space integration over the $r_i+r_f+n_X$ particle final state
yields the total cross section related to the emission of $r_i$ IS and
$r_f$ FS photons [$\beta_e=(1-4m_e^2/s)^{1/2}$, $s=q^2$],
\ba
\sigma_{r_i,r_f}(s) &=& \frac{1}{2s\beta_e}
\int \prod_{a=1}^{r_i}\prod_{b=1}^{r_f}\prod_{c=1}^{n_X}\;
dLips_a^{(i)}\; dLips_b^{(f)}\; dLips_c^{(X)}\nn\\
&& \;\;
(2\pi)^4 \;\delta^{(4)}(q-\sum_{a=1}^{r_i}k^{(i)}_a-\sum_{b=1}^{r_f}k^{(f)}_b
-\sum_{c=1}^{n_X}k^{(X)}_c) \nn\\
&& \;\ol{|{\cal M}^{(r_i,r_f)}|^2} \;,
\label{ph}
\ea
with
\ba
dLips_a^{(i)} &=&
\frac{d^{3} k^{(i)}_a}{(2\pi)^{3} 2 E_a^{(i)}}
,\;
dLips_b^{(f)} =
\frac{d^{3} k^{(f)}_b}{(2\pi)^{3} 2 E_b} ,\; \nn\\
dLips_c^{(X)} &=& \frac{d^3 k^{(X)}_c}{(2\pi)^3 2E^{(X)}_c} .
\ea
For the real photons the same IR regulator (photon mass)
has to be used as for the virtual photons. Hence we are all the time
dealing with IR--regularized expressions.
Although Eq.~(\ref{ph}) includes now virtual and real photon
IS and FS corrections we cannot remove the IR regulator since
we included only $r$ real photons but virtual corrections to
all orders. $\sigma_{r_i,r_f}(s)$ is obviously not
a physical quantity by itself.
Inserting the following identities (with $s_V=q_V^2$),
\ba
1 &=& \int d^4q_V\; \delta^{(4)}(q_V-\sum_{b=1}^{r_f} k_b^{(f)}-
\sum_{c=1}^{n_X} k_c^{(X)}) \nn\\ \quad \rm and \quad && \nn \\
 1 &=& \int ds_V\;\delta(s_V-q_V^2) ,
\ea
into Eq.~(\ref{ph}) yields
\ba
\sigma_{r_i,r_f}(s) &=& \frac{1}{2s\beta_e}
\int ds_V
\int \prod_{a=1}^{r_i}\;dLips_{a}^{(i)}\;\;\frac{d^3q_V}{2q_V^0}\; \nn\\
&\times &\delta^{(4)}(q-q_V-\sum_{a=1}^{r_i} k_a^{(i)})\;\nn\\
&\times &\; E_{\mu\nu}^{(r_i)}(q,q_V,\{k^{(i)}\})
\;{\cal F}^{\mu\nu}_{(r_f)}(q_V) \;,
\label{ISFSfact}
\ea
with the integrated FSR tensor
\ba
{\cal F}^{\mu\nu}_{(r_f)}(q_V) &=&
\int\prod_{b=1}^{r_f}\prod_{c=1}^{n_X}\;dLips_{b}^{(f)}\;dLips_{c}^{(X)}\;
\nn\\
&&(2\pi)^4\;\delta^{(4)}(q_V-\sum_{b=1}^{r_f} k_b^{(f)}-
\sum_{c=1}^{n_X} k_c^{(X)})\;\nn\\
&& \; F^{\mu\nu}_{(r_f)}(q_V,\{k^{(X)}\},\{k^{(f)}\}) \;.
\label{fsrten}
\ea
From Lorentz-covariance follows that the integrated FSR tensor
can be written as a linear combination of the two linear
independent tensors $g^{\mu\nu}$ and $q_V^{\mu}q_V^{\nu}$:
\ba
{\cal F}^{\mu\nu}_{(r_f)}(q_V) &=& A^{(r_f)}(s_V)\;g^{\mu\nu} +
B^{(r_f)}(s_V)\;q_V^{\mu}q_V^{\nu} \nn\\
&=&
A^{(r_f)}(s_V)\left(g^{\mu\nu}-\frac{q_V^{\mu}q_V^{\nu}}{s_V}\right)
\label{decom} \\
&=& \frac{1}{3}\tr\left[{\cal F}^{\mu}_{(r_f)\nu}(q_V)\right]
\left(g^{\mu\nu}-\frac{q_V^{\mu}q_V^{\nu}}{s_V}\right). \nn
\ea
For the second equality in~(\ref{decom}) gauge invariance has been used,
implying the Ward identity $q_{V\mu}{\cal F}^{\mu\nu}_{(r_f)}(q_V)=0$.
Note that for the special case of no ISR photon emission ($r_i=0$)
the related total cross section corresponding to the emission of
$r_f = r$ FSR photons can be written as
\ba
\sigma_{0,r_f}(s) &=& \frac{1}{2s\beta_e}\;E_{\mu\nu}^{(0)}(q)\;
{\cal F}^{\mu\nu}_{(r_f)}(q) \;,
\ea
with the lowest order IS tensor 
\ba
E_{\mu\nu}^{(0)}(q) &=& e^2\;\left(p_{1\mu}p_{2\nu}+p_{2\mu}p_{1\nu}
-\frac{s}{2}\;g^{\mu\nu}\right).
\ea
Taking into account the Ward identity
$q^{\mu} E_{\mu\nu}^{(0)}(q)=0$ and using Eq.~(\ref{decom})
we can write
\ba
\sigma_{0,r_f}(s) &=& \frac{1}{2s\beta_e(s)}\;
\tr[E^{(0)\nu}_{\mu}(q)]\;
\frac{1}{3}\tr\left[{\cal F}^{\mu}_{(r_f)\nu}(q)\right] \nn\\
&=& -\frac{e^2}{3}\;\frac{s+2m_e^2}{2s\beta_e(s)}\;
\tr\left[{\cal F}^{\mu}_{(r_f)\nu}(q)\right]\;.
\label{Anf}
\ea
For the general case including $r_i$ ISR photons
the following Ward identity holds:
\ba
q_V^{\mu}\;E_{\mu\nu}^{(r_i)}(q,q_V,\{k^{(i)}\}) = 0 \;.
\ea
Inserting the expression for ${\cal F}^{\mu\nu}(q_V)$ in
(\ref{decom}) into (\ref{ISFSfact})
we can write the cross section corresponding to the emission
of $r_i$ ISR and $r_f$ FSR photons as
\ba
\sigma_{r_i,r_f}(s) &=& \frac{1}{2s\beta_e}
\int ds_V\;
\frac{1}{3}\tr\left[{\cal F}^{\mu}_{(r_f)\nu}(q_V)\right] \nn\\
&& \int \prod_{a=1}^{r_i} dLips_{a}^{(i)}\;\frac{d^3q_V}{2q_V^{0}}
\delta^{(4)}(q-q_V-\sum_{a=1}^{r_i} k_a^{(i)})\; \nn\\
&&\tr[\;E_{\mu}^{(r_i)\nu}(q,q_V,\{k^{(i)}\})] 
\ea
(note that $\tr\left[{\cal F}^{\mu}_{(r_f)\nu}(q_V)\right]=3A^{(r_f)}(s_V)$
is only a function of $s_V$).
Hence, using Eq.~(\ref{Anf}), we finally arrive at
\ba
\sigma_{r_i,r_f}(s) &=& \int ds_V\; \sigma_{0,r_f}(s_V)
\;\rho_{\rm ini}^{(r_i)}(s,s_V) \;,
\label{conv1}
\ea
with
\ba
\rho_{\rm ini}^{(r_i)}(s,s_V) &=& -\frac{1}{e^2}\frac{1}{s+2m_e^2}
\int \prod_{a=1}^{r_i} dLips_{a}^{(i)}\;\frac{d^3q_V}{2q_V^{0}}\;\nn\\
&&\delta^{(4)}(q-q_V-\sum_{a=1}^{r_i} k_a^{(i)})\;\nn\\
&& \tr[\;E_{\mu}^{(r_i)\nu}(q,q_V,\{k^{(i)}\})] \;.
\ea
[Note that for the case of no ISR photons the above equation directly gives
$\rho_{\rm ini}^{(0)}(s,s_V) = \delta(s-s_V)$ which just projects
out $\sigma_{0,r_f}(s)$ in (\ref{conv1})].
At this point we can ask the question what will be the
expression for the inclusive total cross section $\sigma_{\rm incl}(s)$
with any number of real and virtual IS and FS photons.
Neglecting the IFS contributions
we can immediately write $\sigma_{\rm incl}(s)$ as the incoherent sum of all
ISR$\otimes$FSR contributions:
\ba
\sigma_{\rm incl}(s) &=& \sum_{r_i,r_f=0}^{\infty} \sigma_{r_i,r_f}(s)
\label{inclusive} \\
&=& \int ds_V\;\sigma^{(\gamma)}(s_V)\;\rho_{\rm ini}^{\rm incl}(s,s_V)
+ O(\alpha^2)_{\rm IFS} ,\nn
\ea
with
\ba
\sigma^{(\gamma)}(s_V) &=& \sum_{r_f=0}^{\infty}
\sigma_{0,r_f}(s_V),\;\;
\rho_{\rm ini}^{\rm incl}(s,s_V) = \sum_{r_i=0}^{\infty}
\rho_{\rm ini}^{(r_i)}(s,s_V). \nn\\
\label{inclusive2}
\ea
So in (\ref{inclusive}) we finally expressed the completely inclusive
total cross section
$\sigma_{\rm incl}(s)=\sigma(e^+e^-\to X + \rm photons)$ in a factorized form.
Note that $\sigma^{(\gamma)}(s_V)$ now contains
the real and virtual FS corrections to all orders and
$\rho_{\rm ini}^{\rm incl}(s,s_V)$ contains the
real and virtual IS corrections to all orders.
$\sigma^{(\gamma)}(s)$ and
$\rho_{\rm ini}^{\rm incl}(s,s_V)$ are separately IR finite and
the IR regulator can therefore be removed. This of course has
to be the case since $\sigma_{\rm incl}(s)$ is [up to $O(\alpha^2)$ IFS
effects] a physical observable.

Using the above formulas it is now straightforward to express
$\sigma_{\rm incl}$ in Eq.~(\ref{inclusive}) as a perturbation series
in $\alpha$. For simplicity we will show explicitly the expansion
only to $O(\alpha)$. For this we write the FSR--inclusive cross section as
the expansion
\ba
\sigma^{(\gamma)}(s) &=&
\sum_{r_f=0}^{\infty}\sum_{v_f=0}^{\infty}\sum_{v_f'=0}^{\infty}
\sigma_{(r_f,v_f,v_f')}(s) \;,
\ea
with
\ba
&& \sigma_{(r_f,v_f,v_f')}(s) = \frac{1}{2 s \beta_e(s)}
\int\prod_{b=1}^{r_f}\prod_{c=1}^{n_X}
\;dLips_{b}^{(f)}\;dLips_{c}^{(X)} \nn\\
&& (2\pi)^4\;\delta^{(4)}(q-\sum_{b=1}^{r_f} k_b^{(f)}-
\sum_{c=1}^{n_X} k_c^{(X)})\; \ol{|{\cal M}|}^2_{(r_f,v_f,v_f')} ,
\ea
and
\ba
\ol{|{\cal M}|}^2_{(r_f,v_f,v_f')}
&=& -\frac{e^2}{3}\;(s+2m_e^2)\sum_{s_{(f)}}\sum_{s_{(X)}} \nn\\
&& \biggl[B_{(r_f,v_f)\mu}(q,\{k^{(X)}\},\{k^{(f)}\}) \nn\\
&& B^{*\mu}_{(r_f,v'_f)}(q,\{k^{(X)}\},\{k^{(f)}\}) \biggr] \;.
\ea
Also the inclusive IS radiator we write as an expansion:
\ba
\rho_{\rm ini}^{\rm incl}(s,s_V) &=&
\sum_{r_i=0}^{\infty}\sum_{v_i=0}^{\infty}\sum_{v_i'=0}^{\infty}
\rho_{(r_i,v_i,v_i')}(s,s_V)
\ea
with
\ba
&&\rho_{(r_i,v_i,v_i')}(s,s_V) =
-\frac{1}{4 e^2}\frac{1}{s+2m_e^2}\int\prod_{a=1}^{r_i} dLips_a^{(i)}
\frac{d^3q_V}{2q_V^0} \; \nn\\
&&\delta^{(4)}(q-q_V-\sum_{a=1}^{r_i} k_a^{(i)})
\; A_{\mu}^{(r_i,v_i)}(q,q_V,\{k^{(i)}\}) \nn\\
&& A_{(r_i,v_i')}^{*\mu}(q,q_V,\{k^{(i)}\}) \;.
\ea
The inclusive cross section can then be written as
\ba
\sigma_{\rm incl}(s) &=& \int ds_V \biggl\{\;
\left[\sigma_{(0,0,0)}(s_V)+\sigma_{(1,0,0)}(s_V) \right.\nn\\
&& + \left. \sigma_{(0,1,0)}(s_V)+
\sigma_{(0,0,1)}(s_V)\right] \nn\\
&& \times
\left[\rho_{(0,0,0)}(s,s_V)+\rho_{(1,0,0)}(s,s_V) \right. \nn\\
&&\left.
+
\rho_{(0,1,0)}(s,s_V)+\rho_{(0,0,1)}(s,s_V)\right] \biggr\} + O(\alpha^2)
\nn\\
&=& \sigma^{(\gamma)}(s)
\left[1+\delta_{\rm ini}(s,\Lambda)\right] \nn\\
&+& \int_{s_V^{\rm min}}^{s-2\sqrt{s}\Lambda} ds_V\;
\sigma^{(\gamma)}(s_V)\;\rho_{\rm ini}(s,s_V) + O(\alpha^2), \nn\\
\label{inclusivealpha}
\ea
with
\ba
\sigma^{(\gamma)}(s_V) &=&
\sigma_{(0,0,0)}(s_V)+\sigma_{(1,0,0)}(s_V) \nn\\
&& + \sigma_{(0,1,0)}(s_V)+
\sigma_{(0,0,1)}(s_V) +O(\alpha^2),
\nn\\
\rho_{\rm ini}(s,s_V) &=& \rho_{(1,0,0)}(s,s_V) ,
\nn\\
\delta_{\rm ini}(s,\Lambda) &=&
\int_{s_V^{\rm min}}^s ds_V\;\left[\rho_{(0,1,0)}(s,s_V)+\rho_{(0,0,1)}(s,s_V)
\right] \nn\\
&+& \int_{s-2\sqrt{s}\Lambda}^s ds_V\;\rho_{(1,0,0)}(s,s_V).
\ea
Here $\Lambda$ is the soft photon cut off,
$\sigma_{(0,0,0)}(s)=\sigma^{(0)}(s)$, and
$\rho_{(0,0,0)}(s,s_V)=\delta(s-s_V)$.

The above derivation had many steps, however, it should be stressed
that the most important one to get ISR$\otimes$FSR factorization 
was the use of Lorentz covariance and gauge invariance in
(\ref{decom}).
If kinematical cuts are applied then ISR$\otimes$FSR factorization
breaks down because (\ref{decom}) will not be valid any more.
Perturbatively this occurs already at $O(\alpha)$,
as we show in  Appendix~B.

\section{Scalar and fermionic final state corrections to $\pi^+\pi^-$
production}
\setcounter{equation}{0}

Taking final state corrections up to $O(\alpha)$ into account
the observed pion pair invariant mass distribution can be written as the sum
of an ISR and an FSR contribution:
\ba
\left(\frac{d\sigma}{ds'}\right)_{\rm obs} &=&
\sigma^{(0)}(s')\;\tilde\rho_{\rm ini}(s,s')
+\sigma^{(0)}(s)\;\tilde\rho_{\rm fin}(s,s'). \nn\\
\label{spectral}
\ea
For the analytic expression of the initial state radiator function
$\tilde\rho_{\rm ini}(s,s')$,
including radiative corrections up to leading log $O(\alpha^3)$
and leading initial state $e^+e^-$ pair production contributions,
we refer to (17) in \cite{Hoefer:2001mx}.
Integrating the spectral function in (\ref{spectral})
over $s'$ yields the observed total cross section $\sigma_{\rm obs}(s)$.

The $O(\alpha)$ final state radiator functions, corresponding to hard
photon radiation from scalar particles $\tilde\rho_{\rm fin}(s,s')$
and from fermionic particles $\tilde\rho_{\rm fin}^f(s,s')$, read
($z=s'/s$)
\ba
\tilde\rho_{\rm fin}(s,s') &=&
\frac{1}{s}
\biggl\{-\;\delta(1-z)+ \left[1+\tilde{\delta}_{\rm fin}^{V+S}(s)\right]
 \label{rhofinscal}\\
&\times& B_\pi(s,s')\left[1-z\right]^{B_\pi(s,s')-1}
\quad \;+\; \tilde{\delta}^H_{\rm fin}(s,s')
\biggr\} \;, \nn \\
\tilde\rho_{\rm fin}^f(s,s') &=&
\frac{1}{s}
\biggl\{-\;\delta(1-z)+ \left[1+\tilde{\delta}_{{\rm fin},f}^{V+S}(s)\right]
\label{rhofinferm} \\
&\times & B_\pi(s,s')\left[1-z\right]^{B_\pi(s,s')-1}
\quad \;+\; \tilde{\delta}^H_{{\rm fin},f}(s,s')
\biggr\} \;, \nn
\ea
respectively, with the corresponding hard and soft photon functions
\ba
\tilde{\delta}_{\rm fin}^H(s,s') &=& \frac{2\alpha}{\pi}(1-z)
\frac{\beta_{\pi}(s')}{\beta_{\pi}^3(s)} \;,\label{deltafinH} \\
&& \nn \\
\tilde{\delta}_{{\rm fin},f}^H(s,s') &=&
\frac{\alpha}{\pi}\;\frac{(1-z)\; s}{s+2m_\pi^2}
\;\frac{\beta_\pi(s')}{\beta_\pi(s)}\; \nn \\
&\times & \left[-1+\frac{1}{\beta_\pi(s')}\;
\log\left(\frac{1+\beta_{\pi}(s')}{1-\beta_{\pi}(s')}\right) \right]\;, \\
&& \nn \\
B_{\pi}(s,s') &=&
\frac{2\alpha}{\pi}\frac{s'\beta_{\pi}(s')}{s\beta_{\pi}(s)}\; \nn \\
&\times &
\left[\frac{1+\beta_{\pi}^2(s')}{2\beta_{\pi}(s')}
\log\left(\frac{1+\beta_{\pi}(s')}{1-\beta_{\pi}(s')}\right)-1\right]
\;,\\
&& \nn \\
\tilde{\delta}_{\rm fin}^{V+S}(s) &=& \frac{\alpha}{\pi}
\left\{ \frac{3s-4m_{\pi}^2}{s\beta_{\pi}}
\;\log\left(\frac{1+\beta_{\pi}}{1-\beta_{\pi}}\right)
\right. \nn \\
&-& 2 - 2\log\left(\frac{1-\beta_{\pi}^2}{4}\right) \nn\\
&-&\;
\frac{1+\beta_{\pi}^2}{2\beta_{\pi}}\;\left[
\log\left(\frac{1+\beta_{\pi}}{1-\beta_{\pi}}\right)\biggl[
\log\left(\frac{1+\beta_{\pi}}{2}\right) \right. \nn\\
&+&\; \log(\beta_{\pi})\biggr]
+\log\left(\frac{1+\beta_{\pi}}{2\beta_{\pi}}\right)
\log\left(\frac{1-\beta_{\pi}}{2\beta_{\pi}}\right) \nn\\
&+&\;
2\dilog\left(\frac{2\beta_{\pi}}{1+\beta_{\pi}}\right)
+2\dilog\left(-\frac{1-\beta_{\pi}}{2\beta_{\pi}}\right) \nn \\
&-& \left.\left.
\frac{2}{3}\pi^2 \right]\right\} \;,
\label{deltafinvs} \\
\tilde{\delta}_{{\rm fin},f}^{V+S}(s) &=&
\tilde{\delta}_{\rm fin}^{V+S}(s)
-\frac{\alpha}{\pi}
\frac{1}{2\beta_\pi}\;\log\left(\frac{1+\beta_{\pi}}{1-\beta_{\pi}}\right)
\;.
\label{deltafinvsferm}
\ea
Neglecting soft photon exponentiation, we can write the observed total
cross section as the sum of a soft photon contribution and a hard photon
contribution:
\ba
\sigma_{\rm obs}^{(f)}(s) &=&
\sigma^{(0)}(s)
\left[1+\delta_{\rm ini}(s,\Lambda)+\delta_{\rm fin}^{(f)}
(s,\Lambda)\right] \nn\\
&+&\int_{4m_{\pi}^2}^{s-2\sqrt{s}\Lambda}
ds'\;\rho_{\rm ini}(s,s')\;\sigma^{(0)}(s') \nn \\
&+&\sigma^{(0)}(s)\int_{4m_{\pi}^2}^{s-2\sqrt{s}\Lambda} ds'\;
\rho_{\rm fin}^{(f)}(s,s'),
\label{totalcross}
\ea
where
\ba
\rho_{\rm fin}(s,s')
&=&
\frac{1}{s}\left[\tilde{\delta}_{\rm fin}^H(s,s')+
\frac{B_{\pi}(s,s')}{1-z}\right] \;,\; \\
\rho_{\rm fin}^f(s,s')
&=&
\frac{1}{s}\left[\tilde{\delta}_{{\rm fin},f}^H(s,s')+
\frac{B_{\pi}(s,s')}{1-z}\right] \;, \\
\delta_{\rm fin}(s,\Lambda) &=&
\log\left(\frac{2\Lambda}{\sqrt{s}}\right)\;B_{\pi}(s'=s)
+\tilde{\delta}_{\rm fin}^{V+S}(s)
\label{deltafin}, \\
\delta_{\rm fin}^f(s,\Lambda) &=&
\log\left(\frac{2\Lambda}{\sqrt{s}}\right)\;B_{\pi}(s'=s)
+\tilde{\delta}_{{\rm fin},f}^{V+S}(s)
\label{deltafinferm}.
\ea
Taking now only the leading order contribution to the
total cross section and the
final state corrections into account leads to the
$O(\alpha)$ FSR--inclusive cross section
\ba
\sigma^{(\gamma)(,f)}(s) &=&
\sigma^{(0)}(s)\;\biggl\{ 1+\delta_{\rm fin}^{(f)}
(s,\Lambda) \nn \\
&+&\int_{4m_{\pi}^2}^{s-2\sqrt{s}\Lambda} ds'\;
\rho_{\rm fin}^{(f)}(s,s') \biggr\} \nn\\
&=&
\sigma^{(0)}(s)\;\left[1+\frac{\alpha}{\pi}\;\eta^{(f)}(s) \right] \;.
\label{siggam}
\ea
The analytic $O(\alpha)$ expression for the integrated final state
correction factors read
\beq \begin{array}{l}
\eta(s) =
\frac{1+\bpi^2}{\bpi} \Biggl\{
4 {\rm Li}_2 \left(\frac{1-\bpi}{1+\bpi} \right)+
2 {\rm Li}_2 \left(-\frac{1-\bpi}{1+\bpi} \right) \nn \\
-3 \log \left(\frac{2}{1+\bpi} \right) \;
\log \left(\frac{1+\bpi}{1-\bpi} \right) -
2 \log ( \bpi ) \: \log \left(\frac{1+\bpi}{1-\bpi} \right)
\Biggr\} \\
- 3 \log \left(\frac{4}{1-\bpi^2} \right)
- 4 \log ( \bpi ) \\
+ \frac{1}{\bpi^3} \left[ \frac{5}{4}(1+\bpi^2)^2-2 \right]\:
\log \left(\frac{1+\bpi}{1-\bpi} \right)+
\frac{3}{2} \frac{1+\bpi^2}{\bpi^2}
\end{array} \label{etascalar}
\eeq
for a scalar particle final state and
\beq \begin{array}{l}
\eta^f(s) = \eta(s) +
\frac{1}{s+2 m_\pi^2} \; \frac{1}{2 s \beta_\pi}
\; \biggl[ (s^2+2 m_\pi^4) \log \left(\frac{1+\bpi}{1-\bpi} \right) \\
 - \frac{s \beta_\pi}{2}\;
(5 s -2 m_\pi^2) \biggr]
+ \frac{1}{s^2 \beta_\pi^3} \biggl[ 4 m_\pi^2 (s-m_\pi^2)
\;\log \left(\frac{1+\bpi}{1-\bpi} \right) \nn\\
 - s \beta_\pi (s+2 m_\pi^2) \biggr]
- \frac{1}{2 \beta_\pi} \log \left(\frac{1+\bpi}{1-\bpi} \right)
\end{array} \label{etafermion}
\eeq
for a fermionic final state.

Finally some comments on applying kinematic cuts, leading to a breaking
of ISR$\otimes$FSR factorization (see Appendix~A).
Treating FSR by sQED we can write the spectral
function with C--symmetric kinematic cuts as
\ba
\left(\frac{d\sigma}{ds'}\right)_{\rm obs,cut}
&=& |F_{\pi}^{(0)}(s')|^2\;
\left(\frac{d\sigma}{ds'}\right)_{\rm ini,cut}^{\rm point} \\
&+&
|F_{\pi}^{(0)}(s)|^2\;
\left(\frac{d\sigma}{ds'}\right)_{\rm fin,cut}^{\rm point} +O(\alpha^2),\nn
\ea
where $(d\sigma/ds')^{\rm point}_{\rm ini,cut}$ and
$(d\sigma/ds')^{\rm point}_{\rm fin,cut}$ are the corresponding IS and
FS spectral function for point-like, scalar particles.  Note that
within the considered sQED model we treat the non--perturbative QCD
effects (pion form factor) in a factorized way which means that the
FSR corrections are treated within pure sQED and therefore do not
affect the pion form factor.  By some straightforward manipulations we
can write the total photon--inclusive cross section (\ref{scincl}) for
the given cuts in a similar form as in (\ref{inclusivealpha}).

Let us note that $s_V$ in (\ref{scincl}) is a formal integration
variable since in $\sigma_{\rm cut}^{(\gamma)}(s_V)$ $s_V$ corresponds
to the invariant mass square of the pions including FSR, while in
$\rho_{\rm ini}^{\rm cut}(s,s_V)$ $s_V$ corresponds to the invariant
mass square $s'$ of the pions excluding FSR.

The cut quantities are defined as follows:
\ba
\sigma_{\rm cut}^{(0)}(s) &=& |F_{\pi}^{(0)}(s)|^2\;
\sigma_{\rm cut}^{\rm 0,point}(s)
\label{cut0}
\ea
with
\ba
\sigma_{\rm cut}^{\rm 0,point}(s) &=&
\pi\;\frac{\alpha^2\beta_{\pi}^3\:\cos\Theta_{\pi}^{\rm M}}{2s}\;\left(1-
\frac{1}{3}\;\cos^2\Theta_{\pi}^{\rm M}\right) . \nn\\
\ea
Here $\Theta_\pi^{\rm M}$ is the minimal angle between the pion
momenta and the beam axis allowed by the given cuts.
Similarly, we obtain
\ba
\sigma_{\rm cut}^{(\gamma)}(s) &=& |F_{\pi}^{(\gamma)}(s)|^2\;
\sigma_{\rm cut}^{\rm 0,point}(s)
\label{sigmaDressedCut}
\ea
and
\ba
\rho_{\rm ini}^{\rm (cut)}(s,s') &=&
\frac{1}{\sigma_{\rm (cut)}^{\rm 0,point}(s')}
\;\left(\frac{d\sigma}{ds'}\right)_{\rm ini,(cut)}^{\rm point} ,
\nn\\
\rho_{\rm fin}^{\rm (cut)}(s,s') &=& \frac{1}{
\sigma_{\rm (cut)}^{\rm 0,point}(s)}
\;\left(\frac{d\sigma}{ds'}\right)_{\rm fin,(cut)}^{\rm point} \; .
\label{rhos}
\ea

\end{document}